\documentclass[sn-mathphys]{sn-jnl}% Math and Physical Sciences Reference Style
\usepackage{lineno}

\jyear{2022}%

\theoremstyle{thmstyleone}%
\theoremstyle{thmstyletwo}%

\theoremstyle{thmstylethree}%
\usepackage{siunitx}
\usepackage{graphicx}
\usepackage{subfigure}
\usepackage{threeparttable}
\usepackage{color}
\usepackage{caption}
\usepackage{indentfirst}
\graphicspath{{figures/}}

\begin{document}

%\title[Article Title]{\textit{Insight}-HXMT thermal control state and thermal-deformation influence analysis}
\title[Article Title]{\textit{Insight}-HXMT on-orbit thermal control status and thermal deformation impact analysis}

%%=============================================================%%
\author[1]{\fnm{Ai-Mei} \sur{Zhang}}

\author[1,2]{\fnm{Yi-Fan} \sur{Zhang}}

\author[1]{\fnm{Jin-Yuan} \sur{Liao}}

\author*[1]{\fnm{Yu-Peng} \sur{Xu}}\email{xuyp@ihep.ac.cn}

\author[1]{\fnm{Yu-Sa} \sur{Wang}}

\author[3]{\fnm{Wen-Bo} \sur{Luo}}

\author[3]{\fnm{Yu-Peng} \sur{Zhou}}

\author[3]{\fnm{Zhi-Ying} \sur{Qian}}

\author[1]{\fnm{Xiao-Bo} \sur{Li}}

\author[1]{\fnm{Fang-Jun} \sur{Lu}}

\author[1]{\fnm{Shuang-Nan} \sur{Zhang}}

\author[1]{\fnm{Li-Ming} \sur{Song}}

\author[1]{\fnm{Cong-Zhan} \sur{Liu}}

\author[1]{\fnm{Fan} \sur{Zhang}}

\author[1]{\fnm{Jian-Yin} \sur{Nie}}

\author[1]{\fnm{Juan} \sur{Wang}}

\author[1]{\fnm{Sheng} \sur{Yang}}

\author[1]{\fnm{Tong} \sur{Zhang}}

\author[1]{\fnm{Xiao-Jing} \sur{Liu}}

\author[1]{\fnm{Rui-Jie} \sur{Wang}}

\author[1]{\fnm{Xu-Fang} \sur{Li}}

\author[1]{\fnm{Yi-Fei} \sur{Zhang}}

\author[1]{\fnm{Zheng-Wei} \sur{Li}}

\author[1]{\fnm{Xue-Feng} \sur{Lu}}

\author[1]{\fnm{He} \sur{Xu}}

\author[1]{\fnm{Di} \sur{Wu}}

\affil*[1]{Key Laboratory of Particle Astrophysics, Institute of High Energy Physics, Chinese Academy of Sciences, Beijing 100049, China}
\affil*[2]{University of Chinese Academy of Sciences, Chinese Academy of Sciences, Beijing 100049, China}
\affil*[3]{Beijing Institute of Spacecraft System Engineering, CAST, Beijing 100094, China}

%\affil*[1]{\orgdiv{Key Laboratory of Particle Astrophysics}, \orgname{Institute of High Energy Physics, Chinese Academy of Science}, \orgaddress{\street{19B Yuquan Road, Shijingshan District}, \city{Beijing}, \postcode{100049},  \country{China}}} 
%\affil*[2]{\orgdiv{The General Department}, \orgname{Beijing Institute of Spacecraft System Engineering}, \orgaddress{\street{No.104 Youyi Rd. Haidian District}, \city{Beijing}, \postcode{100094},  \country{China}}}

%\affil[2]{\orgdiv{Department}, \orgname{Organization}, \orgaddress{\street{Street}, \city{City}, \postcode{10587}, \state{State}, \country{Country}}}

%\affil[3]{\orgdiv{Department}, \orgname{Organization}, \orgaddress{\street{Street}, \city{City}, \postcode{610101}, \state{State}, \country{Country}}}

%%==================================%%
%% sample for unstructured abstract %%
%%==================================%%

\abstract
{\textbf{Purpose:} 
The Hard X-ray Modulation Telescope is China’s first X-ray astronomy satellite launched on June 15th, 2017, dubbed \textit{Insight}-HXMT. Active and passive thermal control measures are employed to keep devices at suitable temperatures. In this paper, we analyzed the on-orbit thermal monitoring data of the first 5 years and investigated the effect of thermal deformation on the point spread function (PSF) of the telescopes.\\

\textbf{Methods:} 
We examined the data of the on-orbit temperatures measured using 157 thermistors placed on the collimators, detectors and their support structures and compared the results with the thermal control requirements. The thermal deformation was evaluated by the relative orientation of the two star sensors installed on the main support structure. its effect was estimated with evolution of the PSF obtained with calibration scanning observations of the Crab nebula.\\

\textbf{Conclusion:} The on-orbit temperatures met the thermal control requirements thus far, and the effect of thermal deformation on the PSF was negligible after the on-orbit pointing calibration.\\
}


%%\pacs[JEL Classification]{D8, H51}

%%\pacs[MSC Classification]{35A01, 65L10, 65L12, 65L20, 65L70}
\maketitle

\section{Introduction}\label{sec1}

\begin{figure}[H]%
    \centering
    \includegraphics[width=0.7\textwidth]{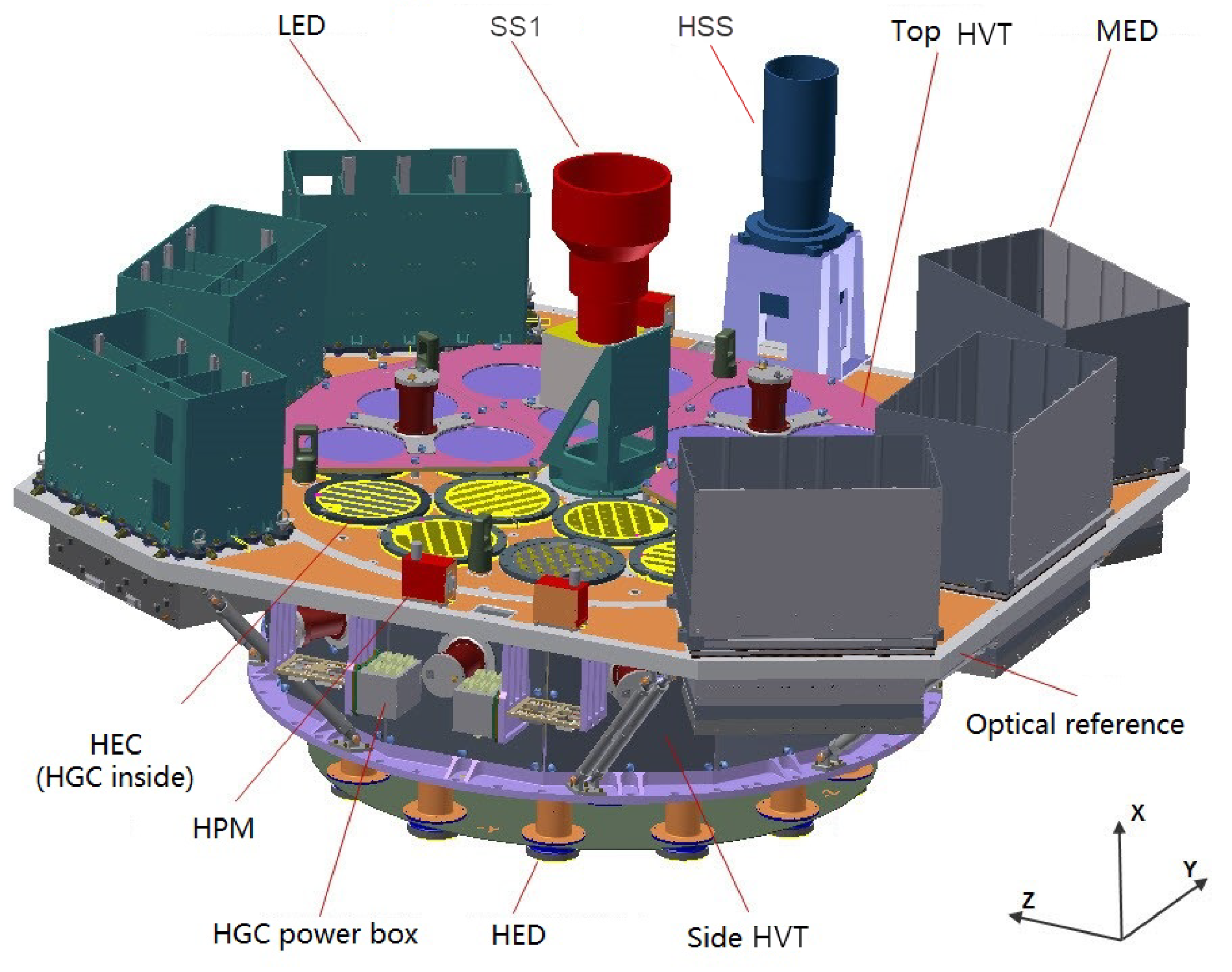}
    \caption{Payloads configuration of \textit{Insight}-HXMT.} \label{fighxmtdistribution}
\end{figure}

\textit{Insight}-HXMT is China’s first X-ray astronomy satellite, launched on June 15th, 2017, which has well operated in orbit for more than five years. \textit{Insight}-HXMT scans the Galactic plane repeatedly, makes pointing observations to neutron stars and stellar mass black holes, and monitors the whole sky continuously in the MeV band. \textit{Insight}-HXMT consists of three main instruments, the high energy X-ray telescope (HE, $20-250$ keV for pointing observations and $0.2-3$ MeV for all-sky monitoring, 5100 $\rm cm^{2}$), the medium energy X-ray telescope (ME, $8-35$ keV, 952 $\rm cm^{2}$), and the low energy X-ray telescope (LE, $1-12$ keV, 384 $\rm cm^{2}$) \cite{zhang2020, Liu2020HE, Cao2020ME, Chen2020LE}. The HE modules are mounted at the center of the main structure, while the LE Detector boxes (LED) and ME Detector boxes (MED) are  mounted on the +Z and -Z side respectively, as shown in Figure \ref{fighxmtdistribution}.  

On one hand, the three telescopes employ different types of detectors to cover a broad energy band of $1-250$ keV. To maximize the performance of the detectors, the three telescopes are designed to operate at quite different temperatures. HE, equipped with NaI(Tl)/CsI(Na) phoswich detectors, is designed to operate at $18\pm2$ {\textcelsius} . ME, equipped with Si-PIN detectors, is designed to operate at $-50$ to  $-5$ {\textcelsius}. LE, equipped with CCD, is designed to operate at $-80$ to $-30$ {\textcelsius}. On the other hand, the axes of HE/ME/LE are expected to be perfectly parallel to each other so as to observe an X-ray source simultaneously with a broad energy band coverage. Therefore, an elaborate thermal control system is designed and implemented to make the detectors work in suitable temperature ranges, and to guarantee the thermal distortion of the reference plane is acceptable at the same time. To achieve this, passive thermal control methods like sun shading, thermal control coatings, heat pipes, thermal insulation technology, etc., and active thermal control methods like heaters, are implemented to ensure the telescope's optimal operating temperatures, and in some cases to maintain the structural stability.

Before launch, we constructed a finite element model of the satellite, which was corrected by the environmental tests. With this model, we can calculate the maximum pointing deviations of the three telescopes caused by thermal distortions assuming the worst temperature conditions that it possibly encounters in orbit. We can also obtain the on-orbit axis direction of the main structure using the two star sensors mounted on the reference plate.

In practice, we usually perform raster scanning of the region containing the Crab nebula from two orthogonal directions to reconstruct the 2-dimensional point spread function (PSF) of each detector unit of the three telescopes\cite{2020JHEAp..25...39N}. Then we can obtain the actual pointing direction of each detector unit.

From the pointing directions obtained by PSFs and the axis direction of the reference plate obtained with the two star sensors, we can further derive the pointing deviations of the detector units of the three telescopes. Comparing the deviations with the results calculated with models, we can get a comprehensive understanding of the thermal deformation effect on the pointing directions of the telescopes.

The influence of the temperature fluctuation on the performance is not included in this paper, which can be found in paper \cite{Ying2023ME} and \cite{Li2023LE}.

\section{Thermal Control Method of \textit{Insight}-HXMT}
 \textit{Insight}-HXMT has an circular orbit with an altitude of 550 km and an inclination angle of 43°. To meet the requirements of the scientific observations, there are three operation modes, the all sky survey mode, the pointing observation mode, and the small region scanning mode. A sun shield is designed to keep the sunlight away and the heat fluxes of the whole payload relatively stable. Normally, the sun shield always faces the sun. However, the thermal radiation environment of \textit{Insight}-HXMT is complicated in all sky survey mode and pointing observation mode, because the Earth's albedo and infrared heat fluxes strongly vary with the attitudes of the satellite. 

Limited by the power supply, the active thermal control method is only implemented on the HE main detector (HED for short) units, which need to operate at a relatively high and precise temperature of $18\pm2\SI{}{\degreeCelsius}$. While passive thermal control methods are implemented on the MED and LED to save power as they operate in a relatively wide temperature range \cite{wang2010}. 
The required operation temperature of HED is $18\pm2\SI{ }{\degreeCelsius}$. The power dissipation of HED is mainly on the readout electronics. %The scintillators are without power. 
Therefore, the HED is covered by MLI and there are two heaters around the scintillators for fine temperature control. For other components of HE, the ranges of operating temperatures are wider and the control method is passive.

The required operating temperature of MED is $-50$ to $\SI{-5}{\degreeCelsius}$. The heat of MED is dissipated by the radiator around the ME instruments, through several heat pipes inside the radiator cooling plate. Since there is no heater for fine control, the temperature of MED depends on the thermal balance of the absorbed heat flow (infrared heat fluxes and earth albedo), the heat dissipation of electronics, and the radiation. The fluctuation is less than 10℃ within one orbit.

The required operating temperature of LED is $-80$ to $\SI{-30}{\degreeCelsius}$. The heat of LED is also dissipated by the radiator around the LE instruments, through several heat pipes. The main material of the LED radiator is the second surface mirror with cerium glass. Its advantage is the high reflection of sunlight and high infrared emissivity, which obtains a very low temperature for LED, about $\SI{-65}{\degreeCelsius}$ in most operation modes. This kind of radiator is not vulnerable to space radiation, which is helpful for the long lifespan requirement of \textit{Insight}-HXMT. The heat pipes of LED are also special because the working temperature of LED is lower than the recommended working temperature range ($-60$ to $\SI{80}{\degreeCelsius}$)\cite{miaoSpacecraftThermalControl2021} of typical ammonia heat pipes. Therefore, the novel low-temperature heat pipe using ethane as working fluid was selected for HXMT. For the temperature requirement of LE electronics and other components, the control method is also passive.%The heat pipes of LED are also special, because the working temperature of LED is lower than  $\SI{-68}{\degreeCelsius}$, while the freezing point of typical ammonia heat pipes is  $\SI{-78}{\degreeCelsius}$. Furthermore, the lower the temperature, the weaker the heat transfer performance. Therefore,  ethane is selected for heat exchange under very low temperatures, with careful verification.% 

\section{on-orbit temperature monitoring data}

\subsection{HE temperature in orbit}\label{sec2}

\begin{figure}[H]%
    \centering
    \includegraphics[width=0.67\textwidth]{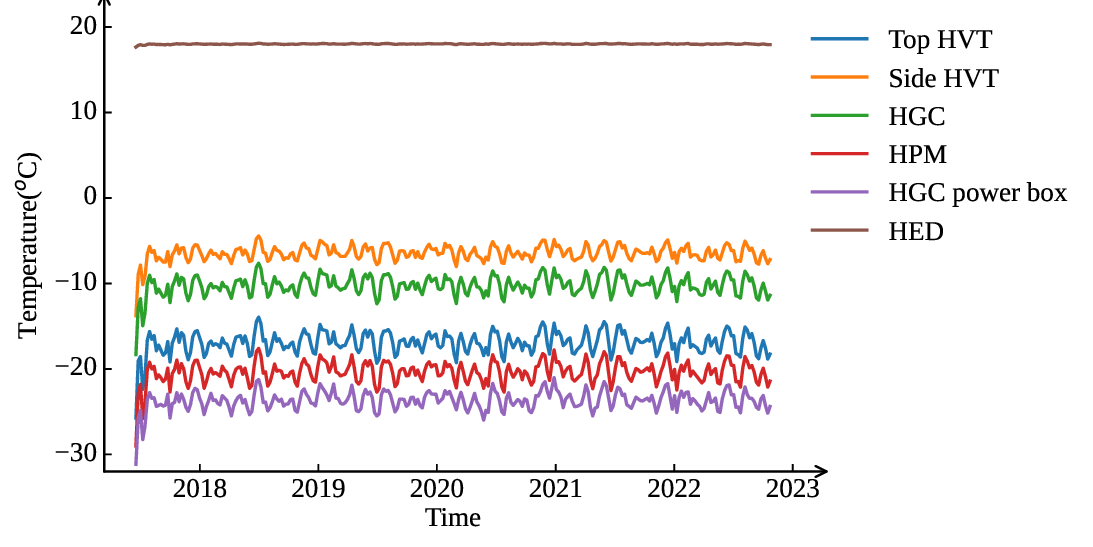}
    \caption{Weekly average temperature of HE subsystems.}
    \footnotesize
    {*}Each curve represents the average temperature of multiple equipment of the same type.
    \label{HEtemp}
\end{figure}

The HE-telescope consists of the HEDs, the high energy collimators (HECs), the auto-gain control detectors (HGCs), the anti-coincidence shield detectors (HVTs), the particle monitors (HPMs), the data processing and control boxes (HEBs), and the power boxes (HEAs). The HE telescopes' weekly average on-orbit temperature, shown in Figure \ref{HEtemp}, exhibits a long term temperature variation trend in the first five years. Each curve represents the average temperature for one type of equipment.

\begin{figure}[H]%
    \centering
    \includegraphics[width=0.67\textwidth]{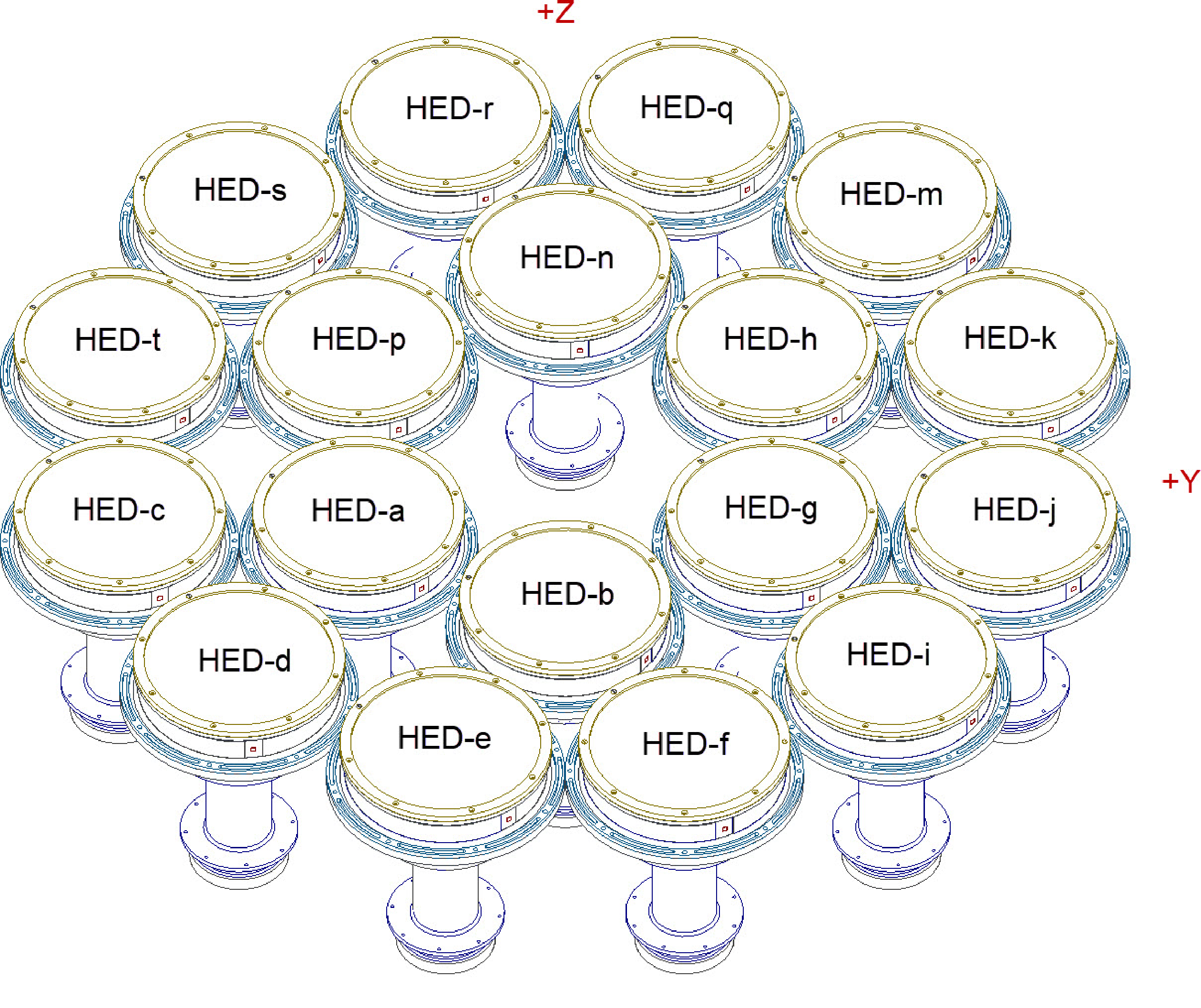}
    \caption{HEDs circular array layout on \textit{Insight}-HXMT.} \label{HEDNO}
\end{figure}

\begin{figure}[H]%
    \centering
    \includegraphics[width=0.45\textwidth]{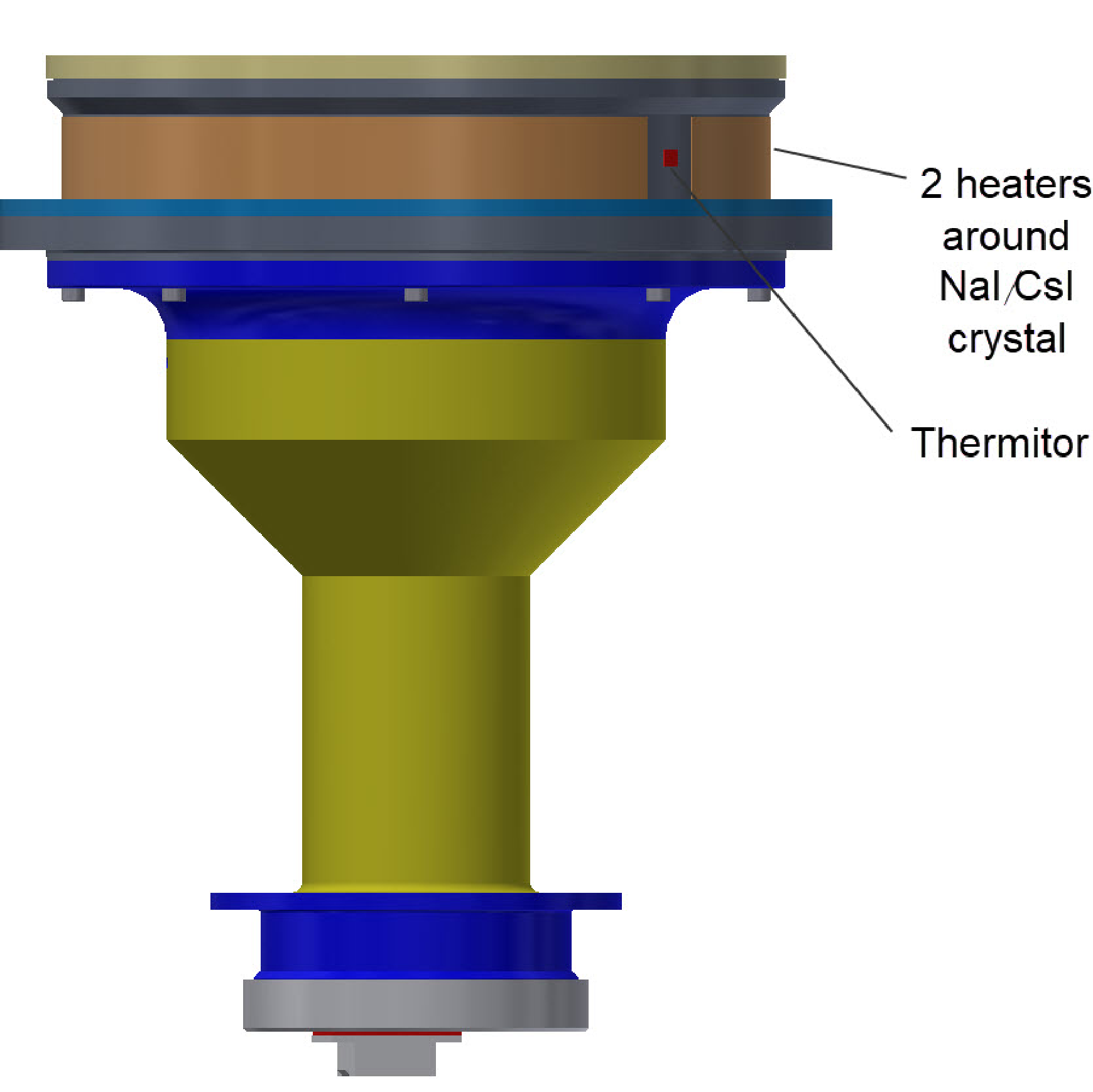}
    \caption{Thermal heaters and thermitors on HED.} 
    \label{HED}
\end{figure}

The 18 HED units are located at the center of the main structure, as shown in Figure \ref{HEDNO}, and are thermally isolated from the surrounding environment.
Each HED has 2 resistive heaters, which almost cover all side-surfaces of the crystal. Two thermistors on both ends of the diameter, as shown in Figure \ref{HED}. The temperature of HEDs are actively controlled at $17.2-18.7\SI{}{\degreeCelsius}$, with covering MLI outside each HED except the crystal surface to save energy. 

\begin{figure}[H]%
    \centering
    \includegraphics[width=0.9\textwidth]{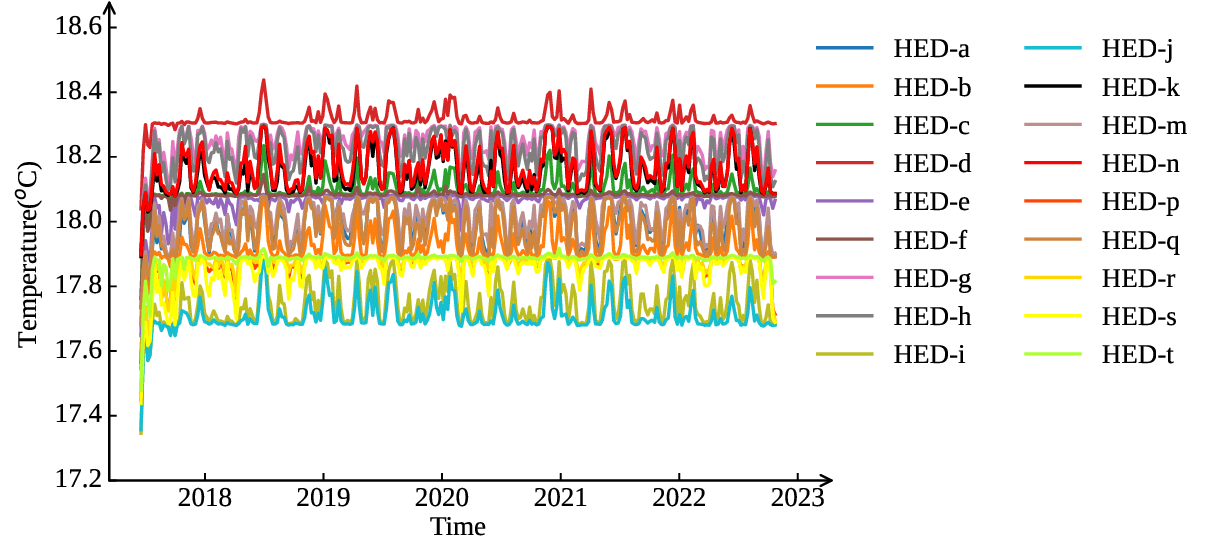}
    \caption{Weekly averaged on-orbit temperature of HEDs.} \label{HEDstemp}
\end{figure}

HEDs' weekly average temperatures, shown in Figure \ref{HEDstemp}, vary slightly, which is similar to the expected state.

\begin{figure}[H]%
    \centering
    \includegraphics[width=0.67\textwidth]{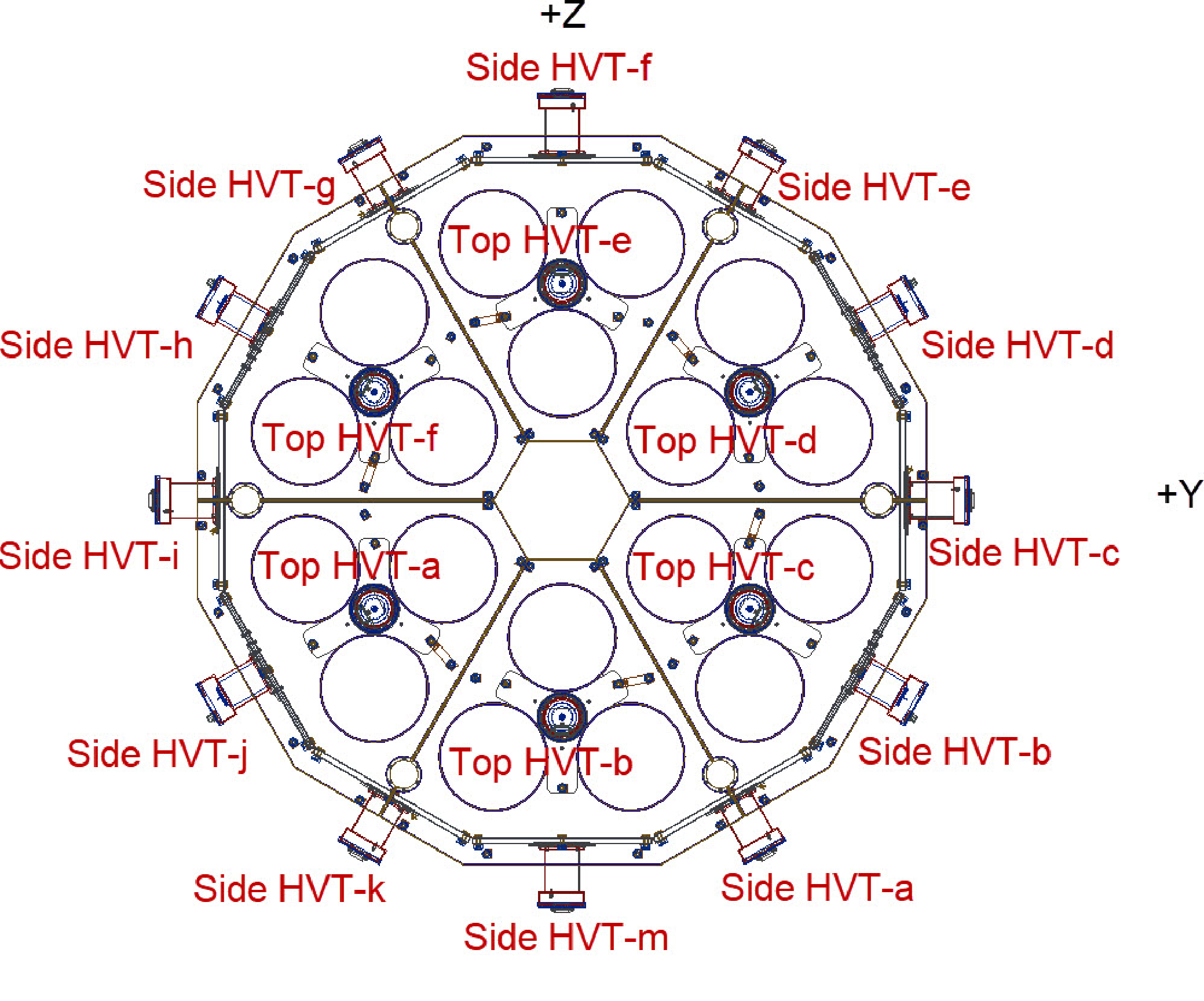}
    \caption{HVTs' layout on \textit{Insight}-HXMT. } \label{HVTNO}
\end{figure}

The 6 sets of top HVT and 12 sets of side HVT, each covered with MLI, are separately integrated on the top and side of the HE collimators with a circular array layout(Figure \ref{HVTNO}). Each HVT has one thermistor on the board close to the electronic box, which monitors the on-orbit temperatures.

\begin{figure}[H]%
    \centering
    \includegraphics[width=0.67\textwidth]{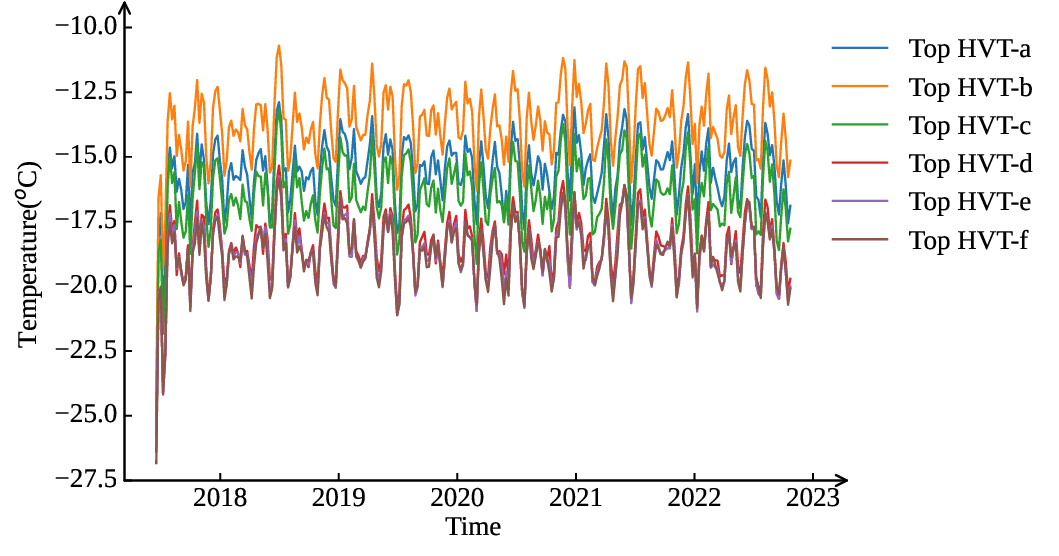}
    \caption{Weekly averaged on-orbit temperatures of the top HVTs.} \label{WeekTopHVT}
\end{figure}

\begin{figure}[H]
    \centering
    \includegraphics[width=0.67\textwidth]{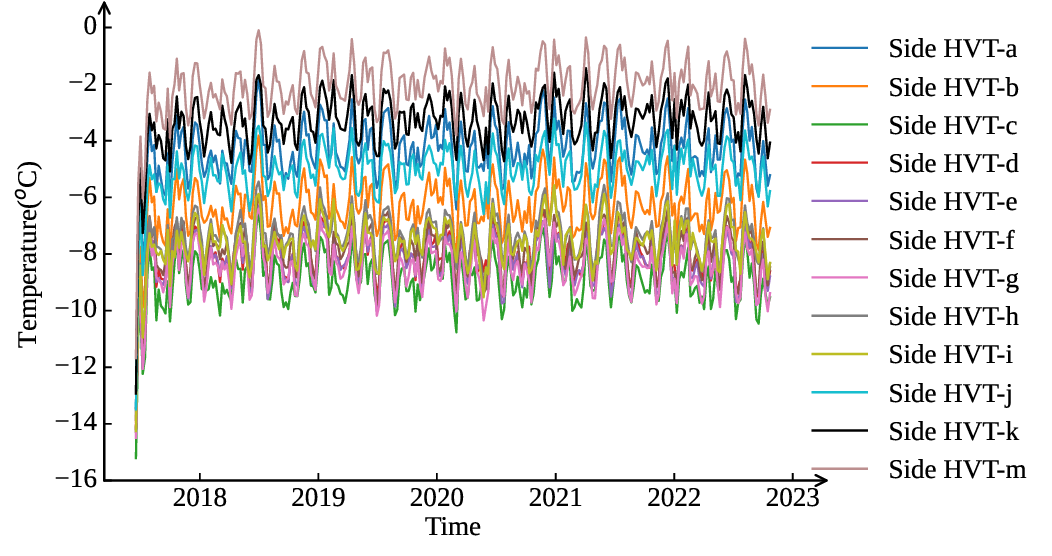}
    \caption{Weekly averaged on-orbit temperatures of the side HVTs.}
    \label{WeeksideHVT}
\end{figure}

The weekly averaged on-orbit temperatures of the top HVTs and side HVTs are separately shown in Figure \ref{WeekTopHVT} and Figure \ref{WeeksideHVT}. Except for the first year in which temperature changes in a larger range corresponding to various observation modes, each HVT’s temperature fluctuating kept in a small range, which means the temperature of single equipment is stable in long term.

\begin{figure}[H]%
    \centering
    \includegraphics[width=0.8\textwidth]{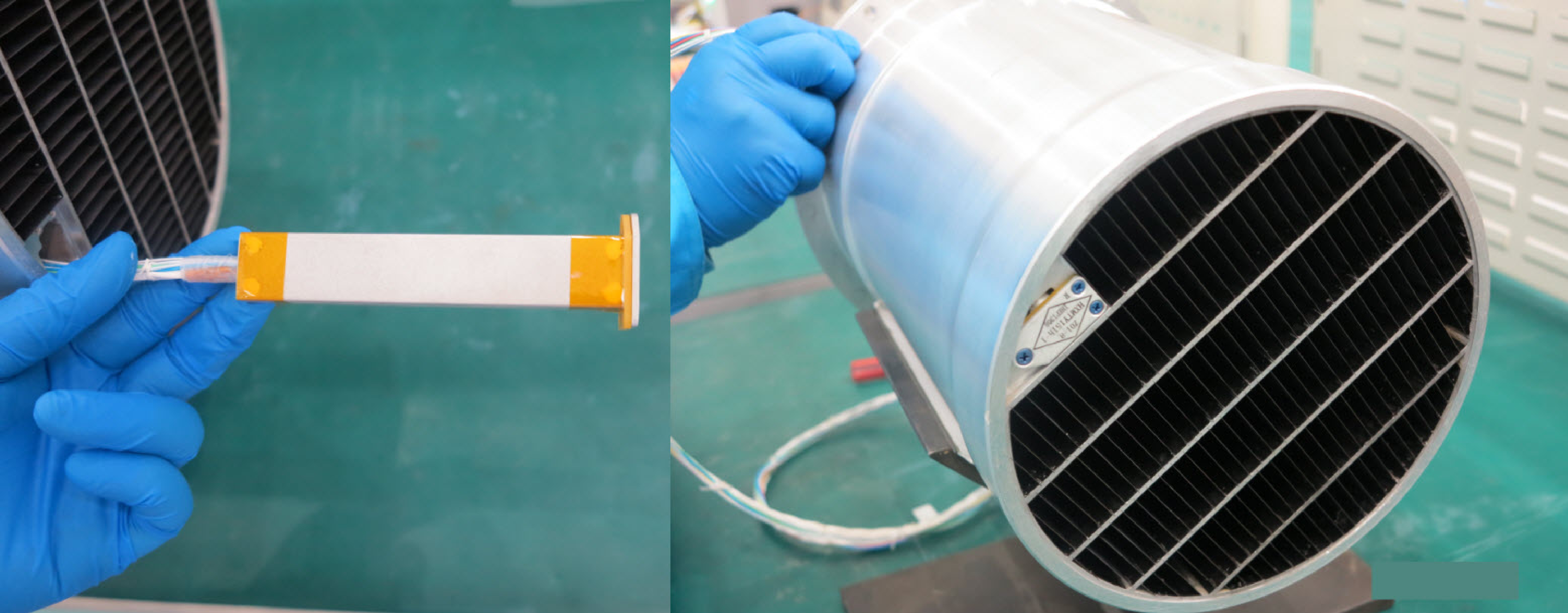}
    \caption{HGC mounted in the HE collimator.} \label{HGC}
\end{figure}

\begin{figure}[H]%
    \centering
    \includegraphics[width=0.67\textwidth]{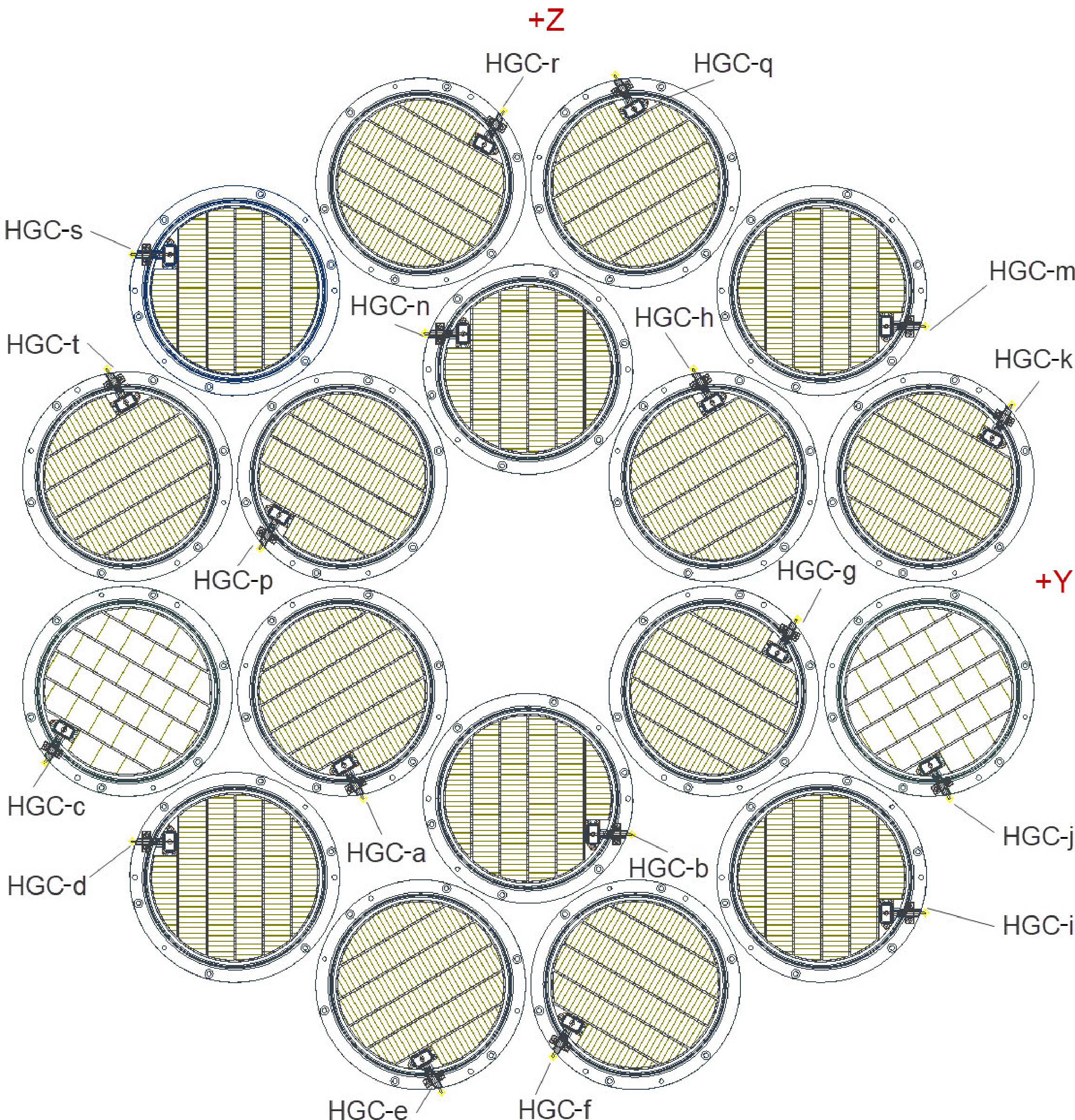}
    \caption{HGC distribution on \textit{Insight}-HXMT.} \label{HGCNO}
\end{figure}

\begin{figure}[H]%
    \centering
    \includegraphics[width=0.8\textwidth]{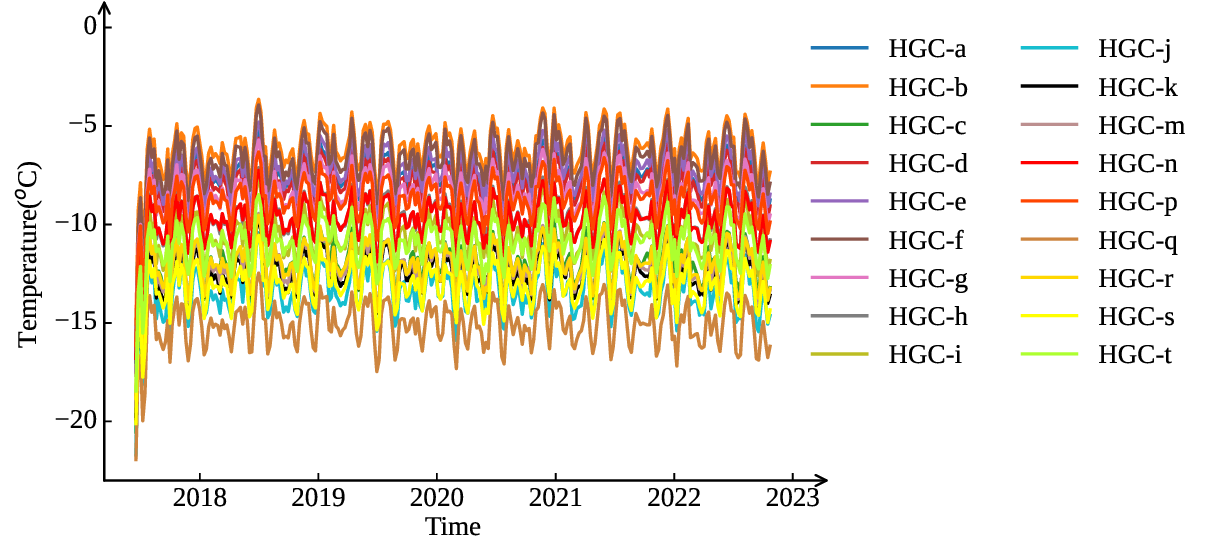}
    \caption{Weekly averaged on-orbit temperatures of the HGC.} 
    \label{WeekHGC}
\end{figure}

Each HGC with one thermistor thermally isolated is mounted in the corresponding HE collimators (Figure \ref{HGC}). Eighteen sets of HE collimators with HCG together are integrated on the top of HEDs with circular array layout as shown in Figure \ref{HGCNO}. The weekly averaged on-orbit temperatures of eighteen HGCs'  (Figure \ref{WeekHGC}) show that the temperature of each set of equipment is stable within a certain equilibrium temperature range.

\begin{figure}[H]
    \centering
    \includegraphics[width=0.7\textwidth]{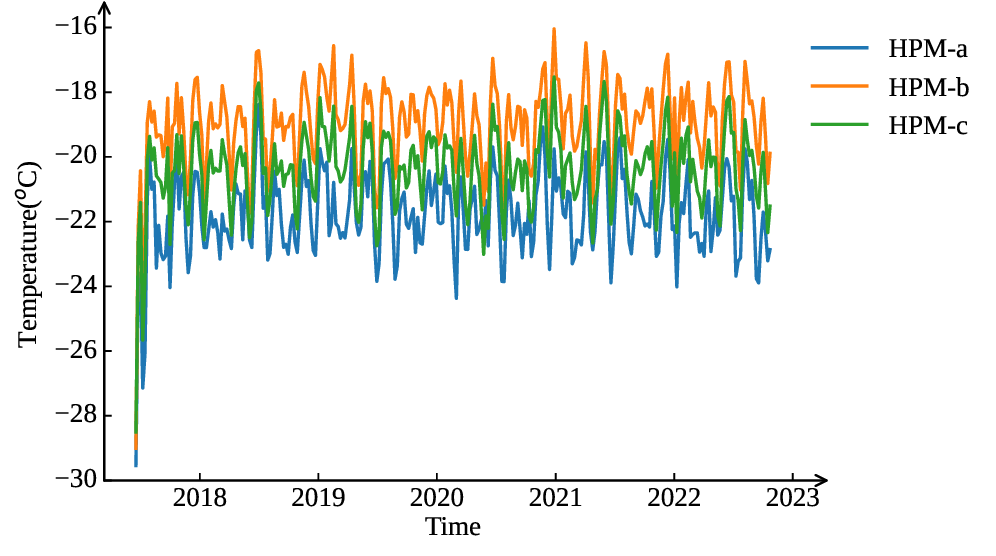}
    \caption{Weekly averaged on-orbit temperatures of the HPMs.}
    \label{WeekHPM}
\end{figure}

\begin{figure}[H]
    \centering
    \includegraphics[width=0.7\textwidth]{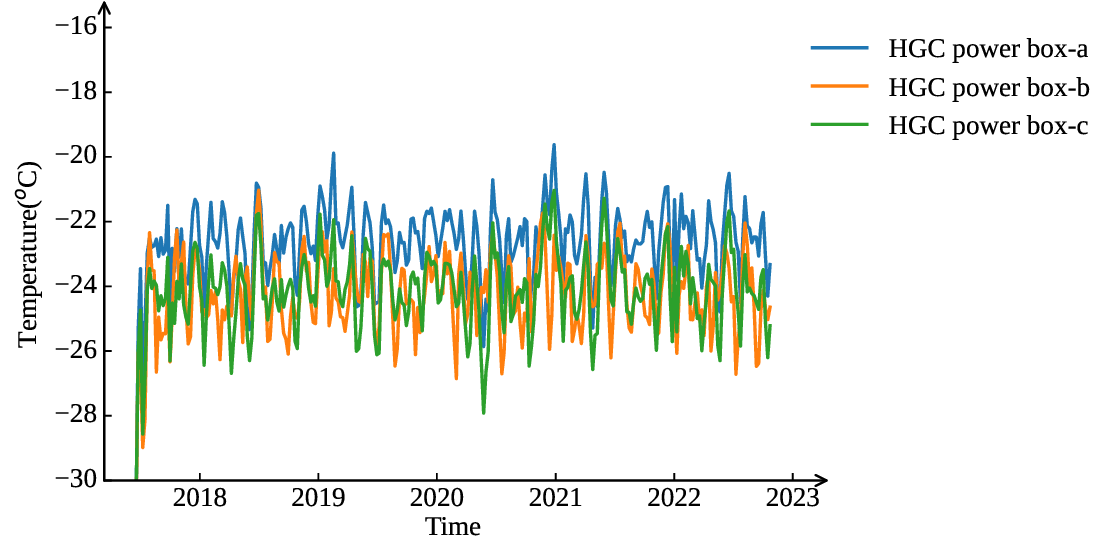}
    \caption{Weekly averaged on-orbit temperatures of the HGC power boxes.}
    \label{WeekHGC_power_box}
\end{figure}

HPMs and HGC power boxes' operating temperature requirement is relatively wide, $-45.0$ to 20.0 {\textcelsius} and $-50.0$ to 20.0 {\textcelsius} respectively. Both on-orbit temperatures match the requirements, shown in Figure \ref{WeekHPM} and Figure \ref{WeekHGC_power_box}.

The on-orbit temperature range of HE devices are listed in Table \ref{tabHE}. All match the thermal control requirements. All of the on-orbit temperatures of the HE units are within the range of predicted temperatures.

The predicted temperatures are analysis results enveloping extreme high-temperature and low-temperature conditions based on the corrected model after the environmental test. During the environmental test, the extreme temperatures could not be achieved, and the temperature envelope range is usually calculated by predicted temperature conditions. The extreme high-temperature condition means that the beta angle reaches 0 degrees (when the satellite has maximum exposure to the sunshine) on the day of the winter solstice (with maximum solar constant). And conversely, extremely low-temperature condition means the beta angle reaches the maximum value of 66.5 degrees at the summer solstice.

\begin{table}[htb]
    \centering
    \caption{Comparison of HE on-orbit temperature with the required and predicted temperature}\label{tabHE}
    \begin{tabular}{|m{60pt}<{\centering}|m{60pt}<{\centering}|m{60pt}<{\centering}|m{60pt}<{\centering}|m{50pt}<{\centering}|}
        \hline
        Equipment name  & Required temperature range ($\SI{}{\degreeCelsius}$) & $\tnote{*}$Predicted temperature  ($\SI{}{\degreeCelsius}$) & on-orbit temperature ($\SI{}{\degreeCelsius}$)\\
        \hline
         HED (with active heating) & $18.0 \pm 2.0$ & 17.4 to 18.7&  17.2 to 18.7 \\
        \hline
        Top HVT & $-48.0$ to 20.0 & $-37.4$  to  $-10.9$ & $-34.6$ to $-9.1$ \\
        \hline
        Side HVT  & $-40.0$ to 20.0 & $-20.6$  to  1.4 & $-19.2$ to 0.5 \\
       \hline
        HGC & $-40$ to 15 &  $-29.8$  to  $-3.8$ &  $-28.7$ to $-3.0$ \\
       \hline
        HPM & $-45.0$ to 20.0 &  $-40.1$  to  $-12.2$ & $-38.1$ to $-14.1$ \\
        \hline
        HGC power box & $-50.0$ to 20.0 & $-40.6$  to  $-14.2$ & $-38.8$ to $-17.5$ \\
        
        \hline
    \end{tabular}
    
% \begin{tablenotes}[H]
% \footnotesize
% \end{tablenotes}
\end{table}

\subsection{MED temperature in orbit}\label{sec3}

\begin{figure}[H]%
    \centering
    \includegraphics[width=0.6\textwidth]{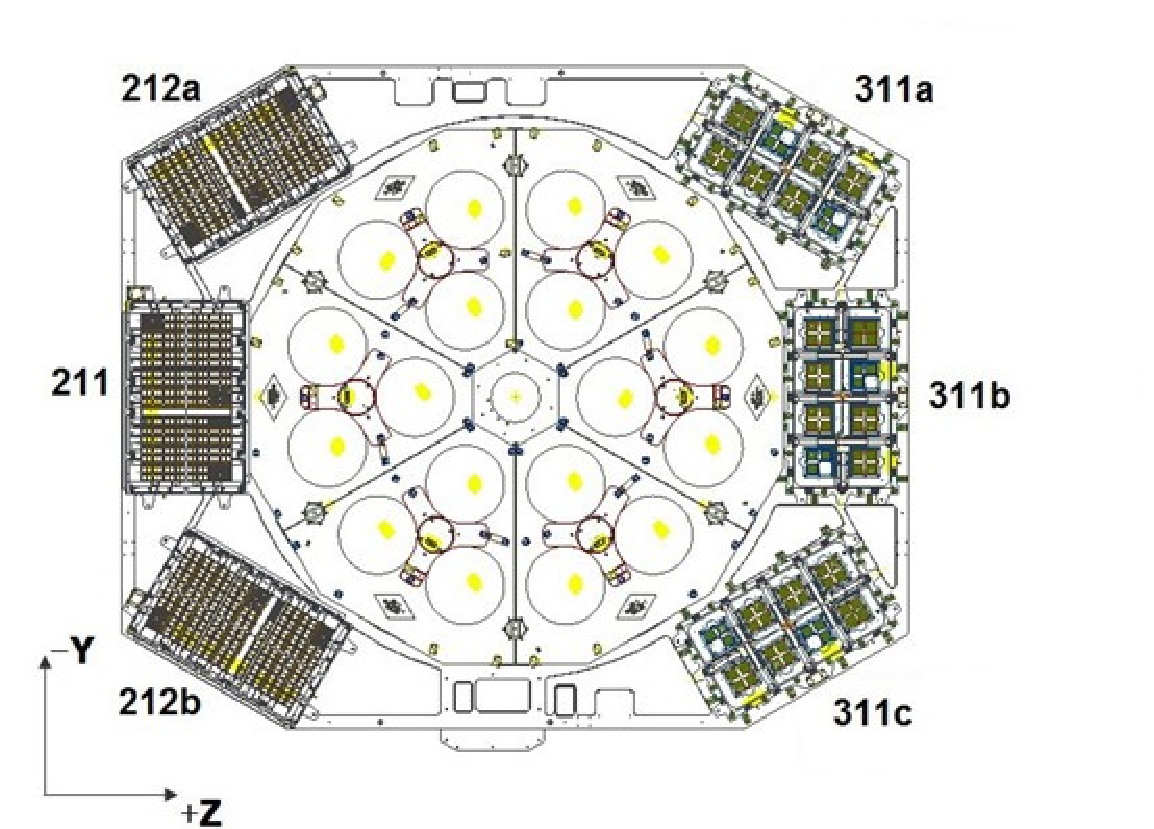}
    \caption{MED and LED layout on \textit{Insight}-HXMT main structure.} \label{figMELEdirtributionn}
\end{figure}

Three sets of MEDs with numbers 211, 212a and 212b are mounted on the -Z side of \textit{Insight}-HXMT optical reference plate, shown in Figure \ref{figMELEdirtributionn}.

\begin{figure}[H]%
    \centering
    \includegraphics[width=0.45\textwidth]{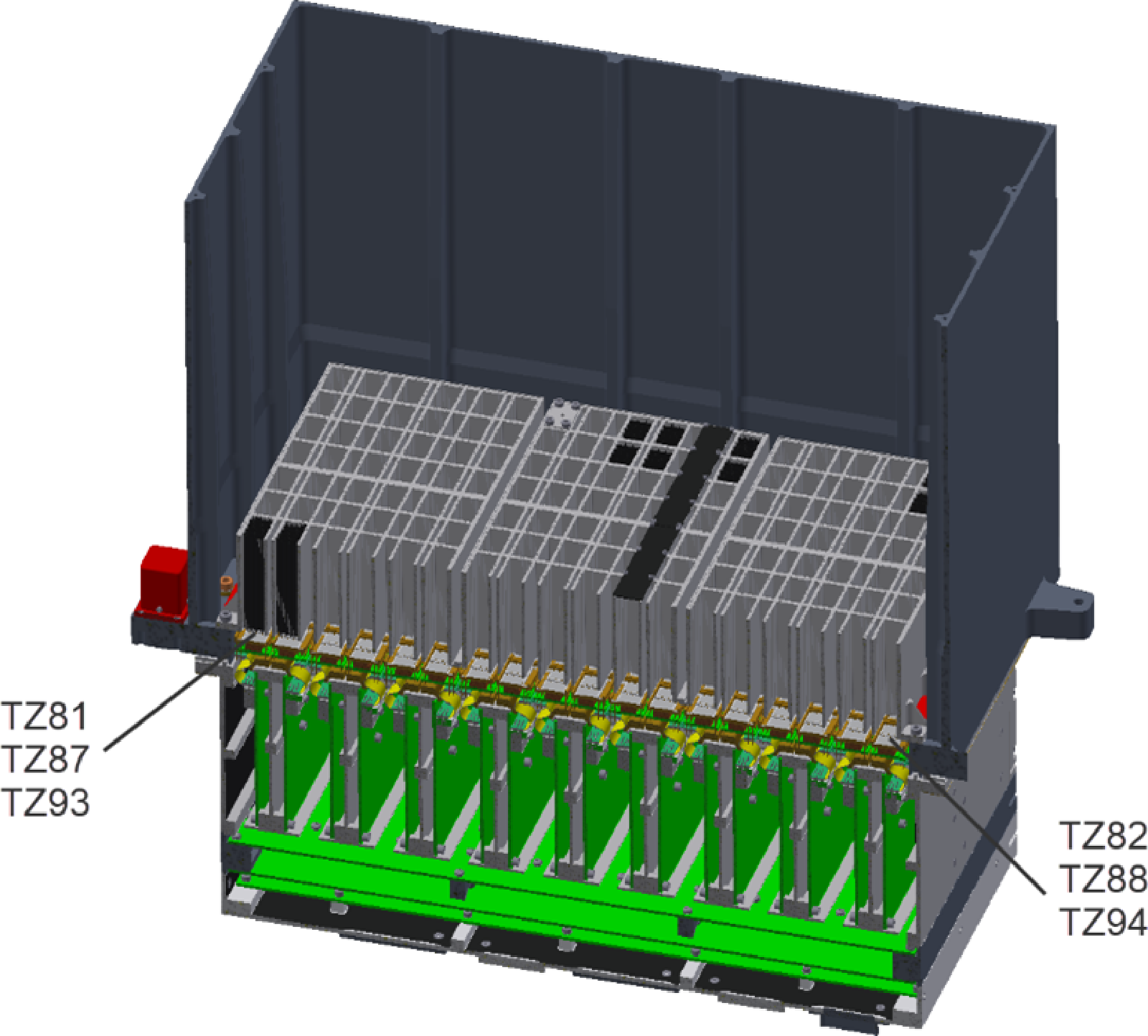}
    \caption{Thermitors' distribution on MED.} \label{figMEdetectorposition}
\end{figure}

\begin{figure}[H]%
    \centering
    \includegraphics[width=0.6\textwidth]{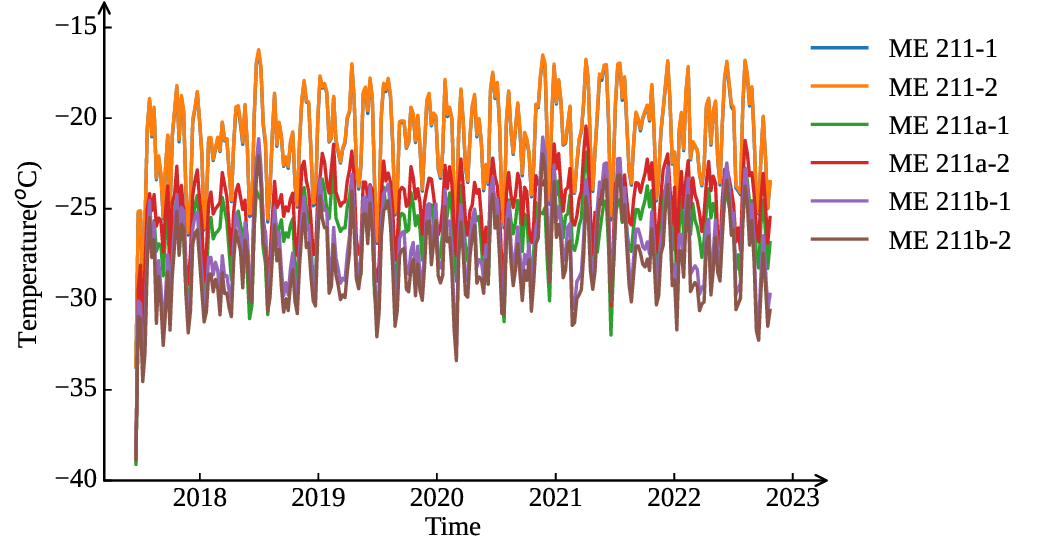}
    \caption{Weekly averaged on-orbit temperatures of the MED detector array.} \label{WeekME}
\end{figure}

The section view of MED is shown in Figure \ref{figMEdetectorposition}, where thermistors TZ81 and TZ82 are set in 211, TZ87 and TZ88 in 212a, TZ93 and TZ94 in 212b. The on-orbit temperature in the first 5 years is shown in Figure \ref{WeekME}. All sky survey observations were carried out for eight periods$\footnote{The eight periods were: 2017-06-15 T22:10:43 to 2017-06-16 T21:48:54, 2017-06-17 T20:57:20 to 2017-06-18 T17:38:04, 2017-07-05 T23:43:24 to 2017-07-08 T01:08:48, 2017-07-10 T00:45:34 to 2017-07-11 T00:43:36, 2017-07-14 T00:11:45 to 2017-07-15 T00:10:08, 2017-09-29 T00:00:05 to 2017-09-29 T17:41:53, 2017-10-25 T02:53:16 to 2017-10-26 T01:19:48, and 2018-07-14 T00:29:20 to 2018-07-15 T02:07:53.}$. Except for these periods, the other observations have been all controlled according to the requirement of the pointing observation mode. During the conversion process of different observation sources and the conversion process between pointing observation and small region scanning modes, a short period of low temperature always occur before the temperature reaches a new equilibrium, which period should not be recorded as the thermal control assessment stage. But the duration before the temperature equilibrium is not definite due to the complexity and diversity of observation attitude changes, so it's very difficult to eliminate from a large amount of data, so this period of low temperature is included in the statistics of the on-orbit temperature range of MED (Table \ref{tabME}) and LED (Table\ref{tabLE}), that’s the reason why the minimum temperature lower than the predicted temperature range. But the in-orbit temperature range meets the temperature control requirements.

\begin{table}[htb]
    \centering
    \caption{Comparison of MED on-orbit temperature with the required and predicted temperature}\label{tabME}
    \begin{tabular}{|m{50pt}<{\centering}|m{80pt}<{\centering}|m{80pt}<{\centering}|m{80pt}<{\centering}|}
        \hline
        Equipment name  & Required temperature range ($\SI{}{\degreeCelsius}$)  & Predicted temperature ($\SI{}{\degreeCelsius}$)&  on-orbit temperature ($\SI{}{\degreeCelsius}$)\\
        \hline
         Detector array  & Detector array: \newline $-50.0$ to $-10.0$ (All sky survey mode) \newline $-50.0$ to $-5.0$ \newline(Pointing observation mode) \newline Temperature uniformity in single detector box: $<5.0$  &  $-48.6$ to $-25.3$ \newline(All sky survey mode) \newline $-20.5$ to $-13.4$\newline (Pointing observation mode) \newline Temperature uniformity in single detector box: $<1.8$ &  $-46.3$ to $-18.6$ \newline(All sky survey mode) \newline $-46.0$ to $-11.4$\newline (Pointing observation mode) \newline  Temperature uniformity in single detector box: $<2.1$\\
        \hline
        Detector box & $-40.0$ to 25.0  & $-32.6$  to  $-6.2$ & $-37.2$ to $-2.8$ \\
        \hline
    \end{tabular}
\end{table}

\subsection{LED temperature in orbit}\label{sec4}

\begin{figure}[H]%
    \centering
    \includegraphics[width=0.5\textwidth]{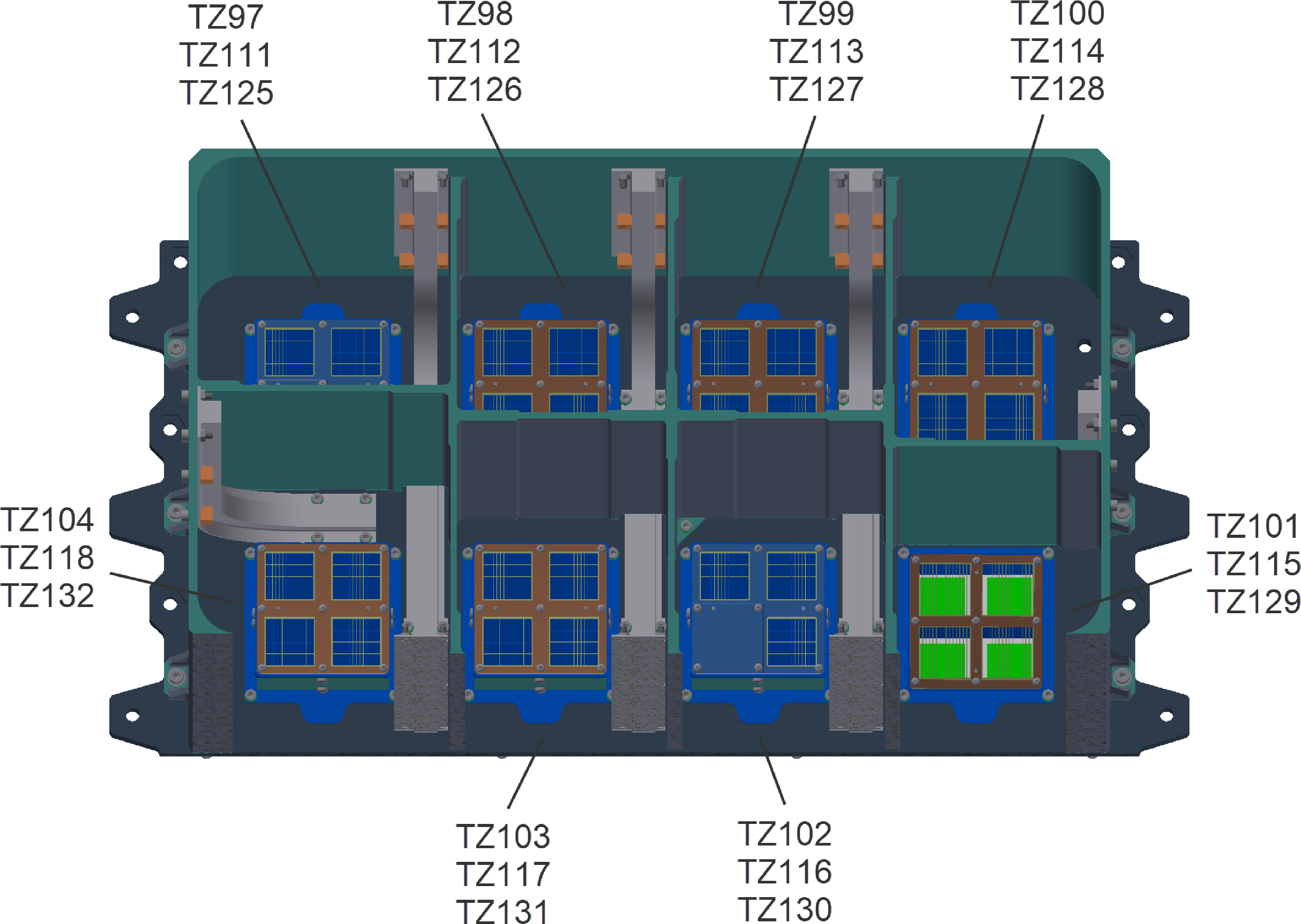}
    \caption{Thermitors' distribution on LED detector array.eps} \label{figLEdetectorposition}
\end{figure}

Three LEDs with numbers 311a, 311b, and 311c are mounted on the -Z side of \textit{Insight}-HXMT main structure, shown in Figure \ref{figMELEdirtributionn}. The thermistors in MED are shown in Figure \ref{figLEdetectorposition}, where TZ97$\sim$TZ104 are in 311a, TZ111$\sim$TZ118 in 311b, and TZ125$\sim$TZ132 in 311c. The observation mode period corresponding to the thermal control requirements is consistent with that of MED.

\begin{figure}[H]%
    \centering
    \includegraphics[width=0.6\textwidth]{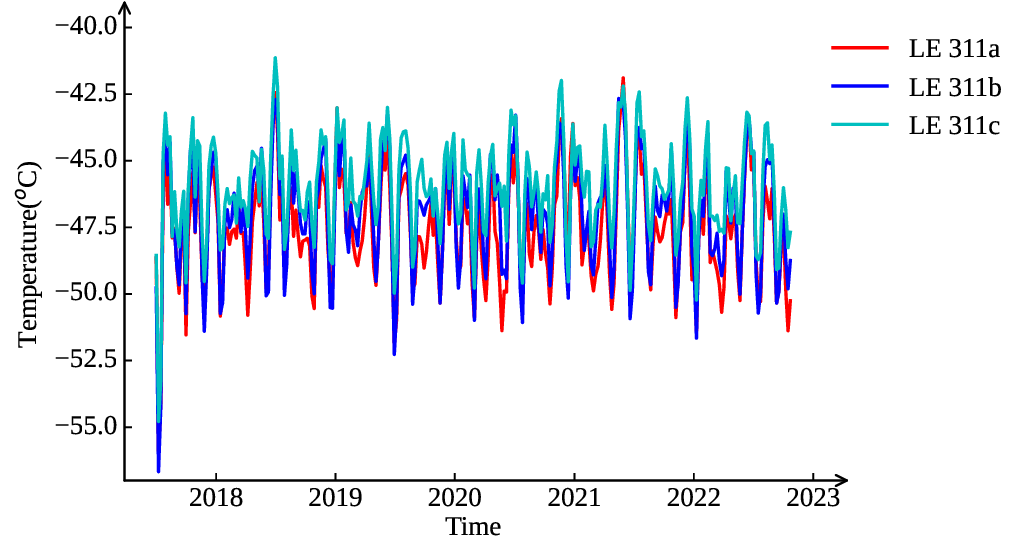}
    \caption{Weekly averaged on-orbit temperatures of the LED detector array.} \label{WeekLE}
\end{figure}

We averaged the temperature values in each set of LED to present long-term temperature variation trends of the three sets of LED, shown in Figure \ref{WeekLE}. We can find that the temperature variation trends of the LEDs are very consistent to each other. 

\begin{figure}[H]%
    \centering
    \includegraphics[width=0.5\textwidth]{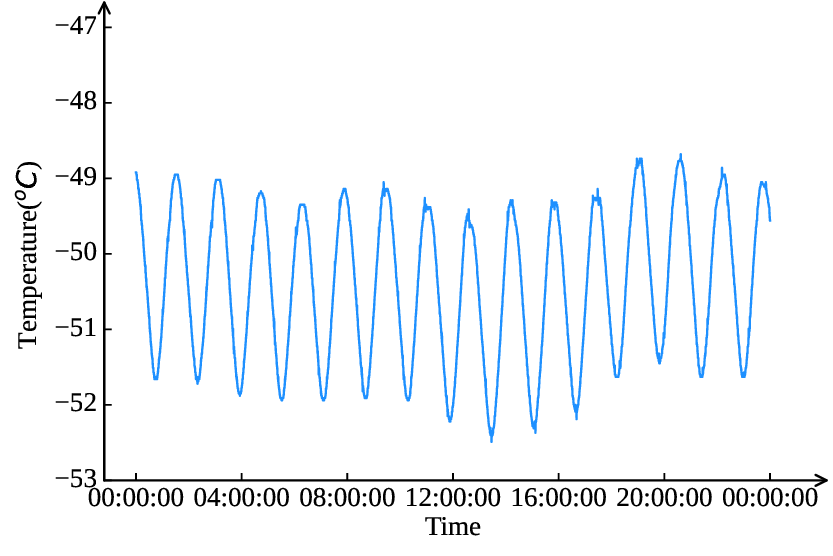}
    \caption{Short-term temperature variations in the low-temperature working condition (June 1st, 2018).} \label{LEDatempwinter}
    \end{figure}

\begin{figure}[H]%
    \centering
    \includegraphics[width=0.5\textwidth]{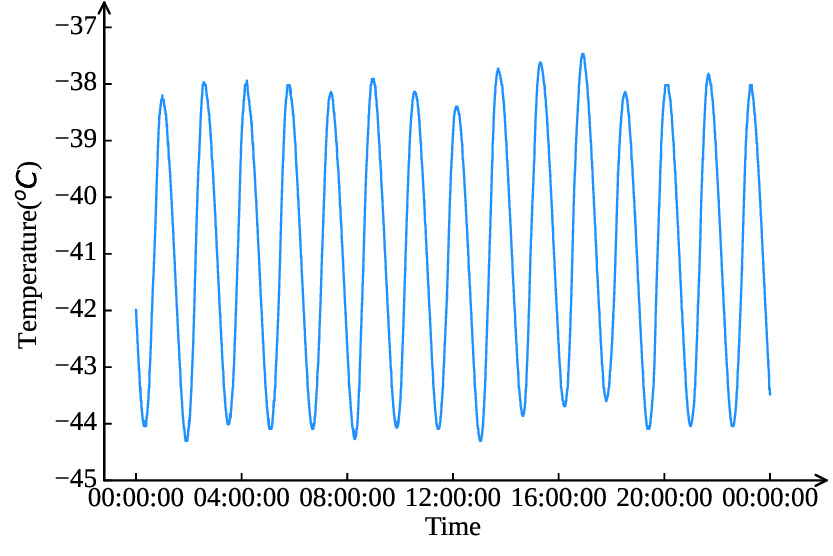}
    \caption{Short-term temperature variations in the high-temperature working condition (January 5th, 2019).} \label{LEDatempsummer}
    \end{figure}

Satellite temperature depends on three factors: beta angle(angle between orbit plane and solar vector), solar constant(related to the period in one year), and the  attitude of the satellite. 
During pointing observations, the attitude has a greater impact on the temperature. 
Taking the short-term on-orbit temperature of LED-311a as a representative example, the temperature level mainly depends on the attitude determined by the target source. 
The lowest and highest working temperature occurred near the summer solstice in 2018 and winter solstice in 2019. 
When we want to find out the lowest temperature near the summer solstice and the highest temperature near the winter solstice in 2018, they separately appeared on June 1st, 2018 (Figure \ref{LEDatempwinter}) and January 5th, 2019 (Figure \ref{LEDatempsummer}). 
Variation amplitude is less than 7 degrees in high-temperature condition and 4 degrees in low-temperature condition in one orbital period. The short-term temperature changes relatively smoothly.

The on-orbit temperature of LED, listed in Table \ref{tabLE}, match the temperature-control requirement well according to observation modes.

\begin{table}[htbp]
    \centering
    \caption{Comparison of LE on-orbit temperature with the required and predicted temperature}\label{tabLE}
    \begin{tabular}{|m{50pt}<{\centering}|m{80pt}<{\centering}|m{80pt}<{\centering}|m{80pt}<{\centering}|}
        \hline
        Equipment name  & Required temperature range ($\SI{}{\degreeCelsius}$)  & Predicted temperature ($\SI{}{\degreeCelsius}$)&  on-orbit temperature ($\SI{}{\degreeCelsius}$)\\
        \hline
         LE detector array & Detector array: \newline $-80.0$ to $-42.0$ \newline (all sky survey mode) \newline $-80.0$ to $-30.0$ \newline(pointing observation mode) \newline Temperature uniformity in single detector box: $<5.0$   & $-75.7$ to $-50.8$ \newline (all sky survey mode) \newline $-41.7$ to $-38.7$ \newline (pointing observation mode) \newline Temperature uniformity in single detector box: $<0.8$ & $-76.1$ to $-44.9$ \newline (all sky survey mode) $-56.6$ to $-36.1$ \newline(pointing observation mode) 
 \newline Temperature uniformity in single detector box:$<1.5$\\
        \hline
        LE detector box  & $-40.0$ to 25.0 &  $-32.6$ to $-13.5$ & $-38.1$ to $-12.5$\\
        \hline
    \end{tabular}
\end{table}

%\section{Pointing analysis}
\section{Thermal deformation analysis}

%\subsection{Thermal deformation analysis } \label{sec5}
\subsection{Simulation study of thermal deformation with predicted temperature} \label{sec5}
\begin{figure}[H]%
    \centering
    \includegraphics[width=0.6\textwidth]{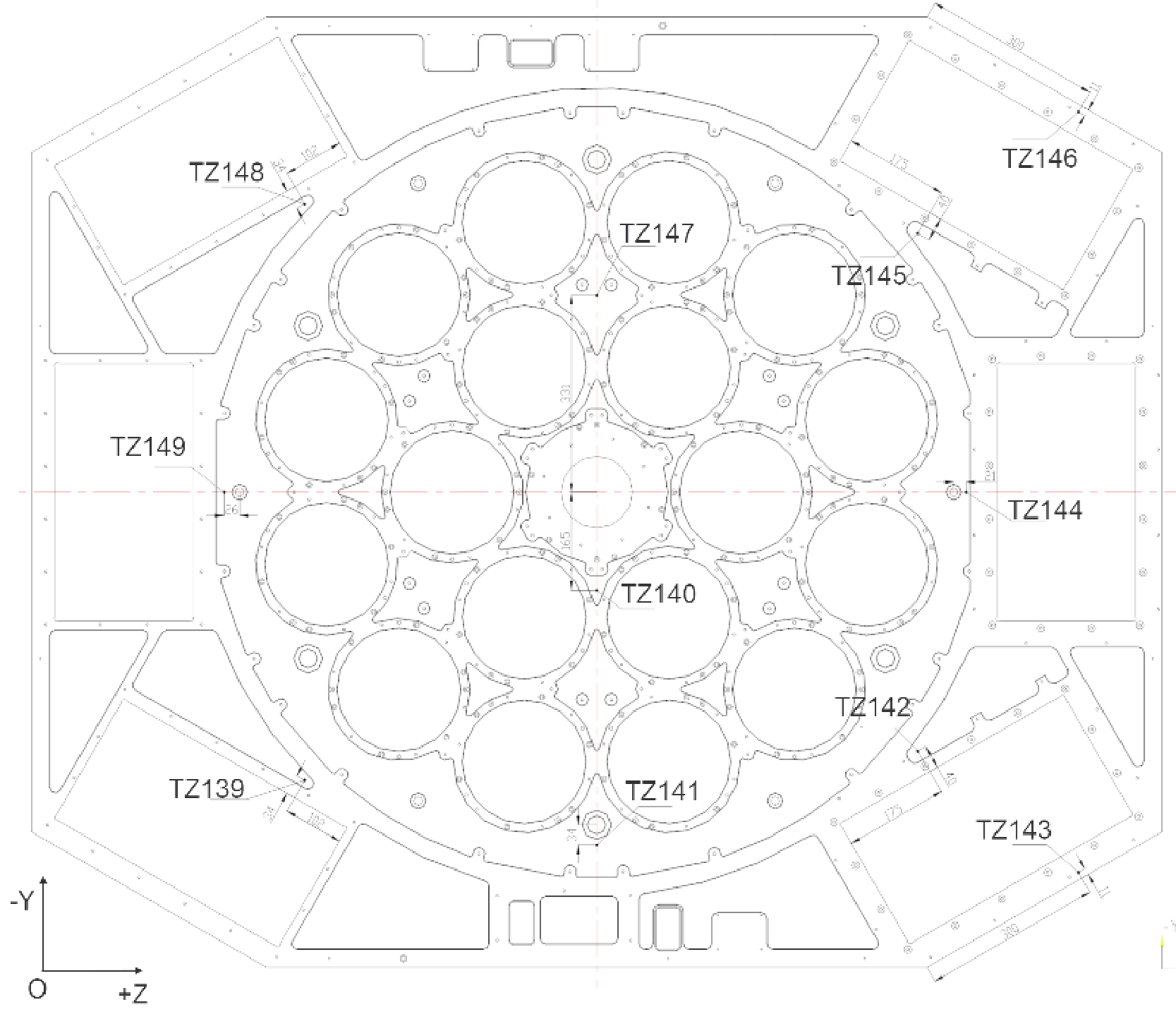}
    \caption{Thermitors' distribution on the pricition reference surface.} \label{figPlatedistribution}
\end{figure}

\begin{figure}[H]%
    \centering
    \includegraphics[width=0.6\textwidth]{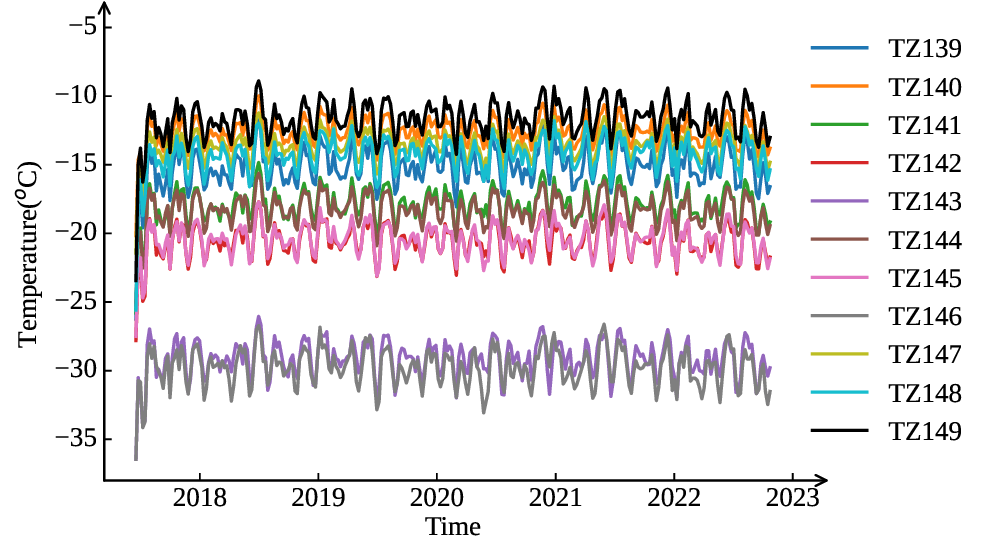}
    \caption{Weekly averaged temperatures of the precision reference surface on which integrates telescopes.} \label{Week_Optical_reference_plate}
\end{figure}

Thermistors set on the optical reference plate are shown in Figure \ref{figPlatedistribution}, monitored temperatures are shown in Figure \ref{Week_Optical_reference_plate}. The on-orbit temperature is very consistent with the data from the thermal balance test. Both TZ143 and TZ146 have lower temperatures since they are near to the radiant surface at the distal edge of the main structure. 

\begin{figure}[H]%
    \centering
    \includegraphics[width=0.45\textwidth]{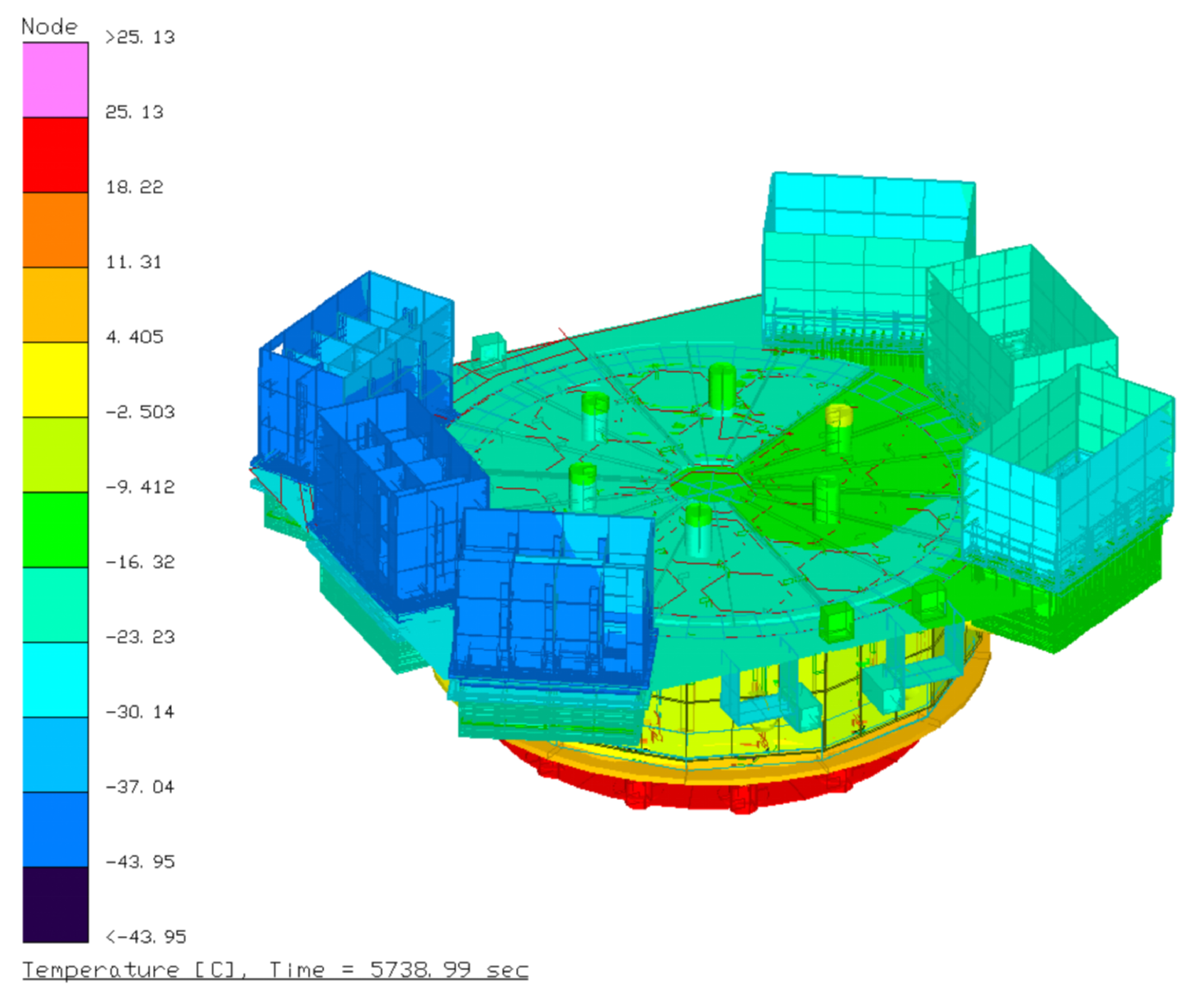}
    \caption{Temperature gradient distribution in high temperature working condition.} \label{Temperaturecloudpicture}
\end{figure}

\begin{figure}[H]%
    \centering
    \includegraphics[width=0.45\textwidth]{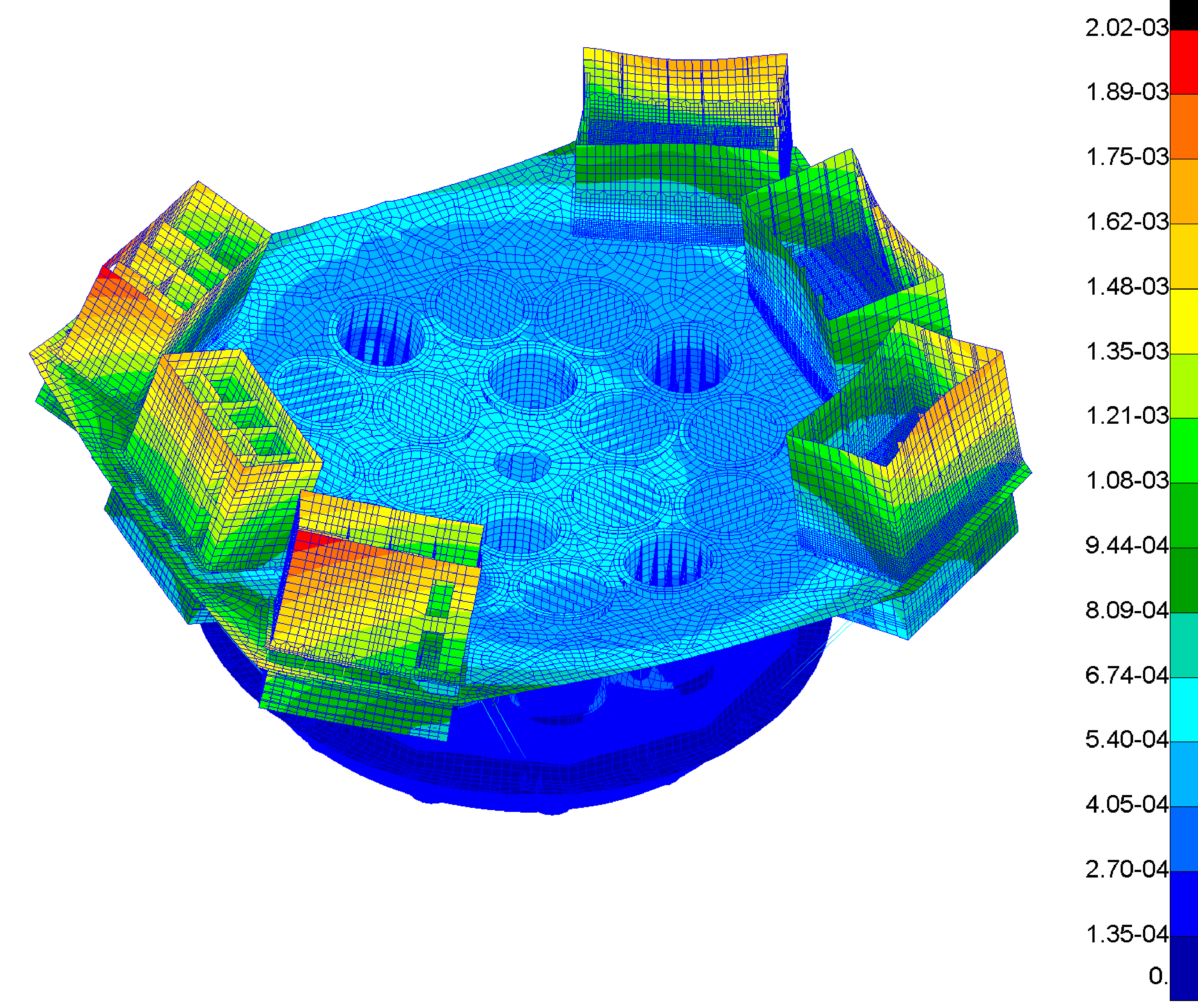}
    \caption{Thermal deformation of \textit{Insight}-HXMT in high temperature working condition.} \label{FEADeformationresult}
\end{figure}

\begin{figure}[H]%
    \centering
    \includegraphics[width=0.45\textwidth]{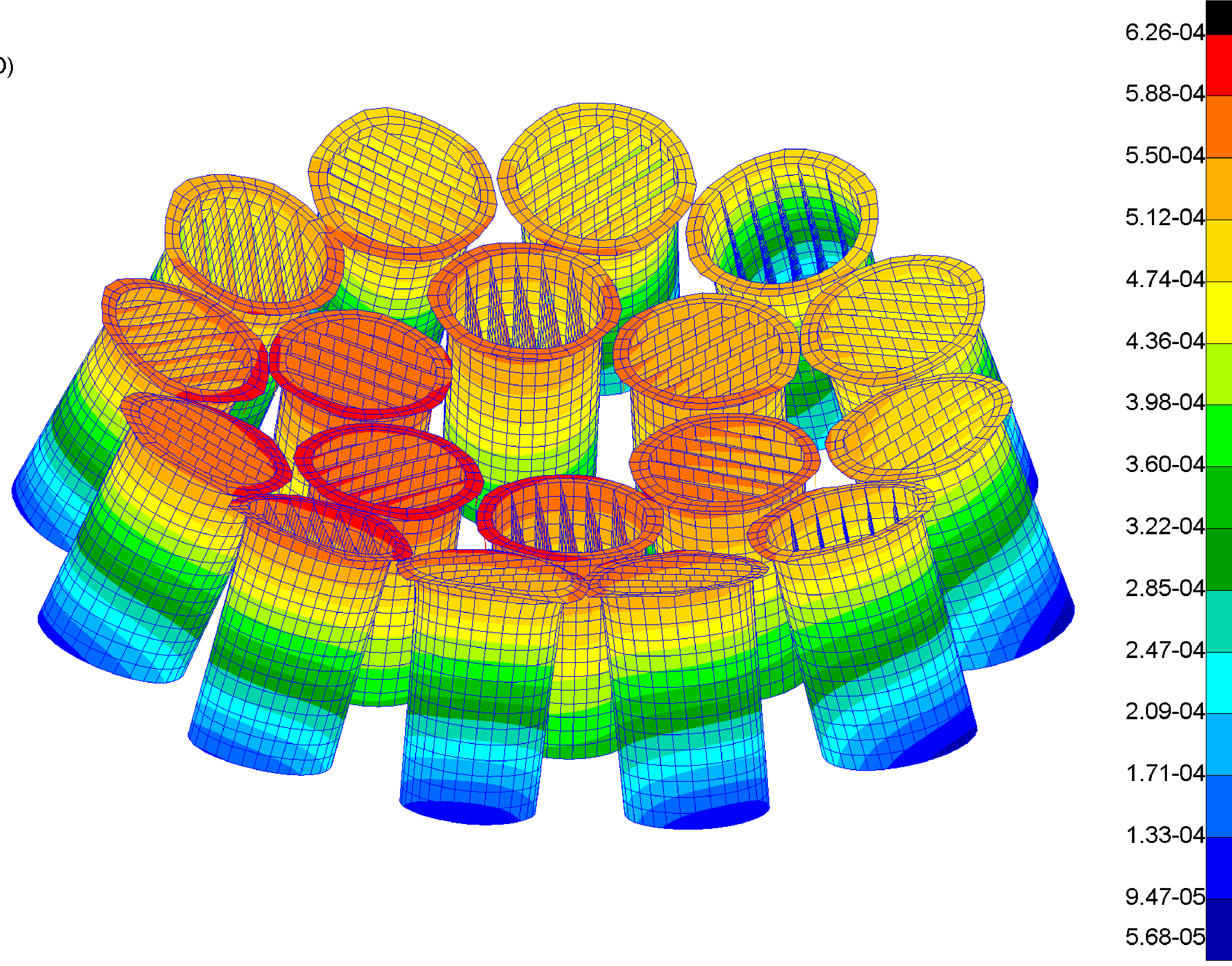}
    \caption{Thermal-deformation of HE collimators in high temperature working condition.} \label{HEDeformationresult}
\end{figure}

Both changes in overall temperature and temperature gradients can cause deformation of the optical reference plate, leading the devices to point in the wrong direction simultaneously. Because the on-orbit resource is limited, only 11 thermitors are set on the reference surface, which are not enough to interpolate and generate the entire temperature field. To predict the on-orbit thermal deformation, we used the temperature field obtained from the thermal balance test to correct the finite element model, and the  temperature gradient result is shown Figure \ref{Temperaturecloudpicture}. Then we simulated the pointing axis of each telescope under extreme thermal conditions according to seasons and the satellite attitudes. A typical optical reference plate deformation map is shown in Figure \ref{FEADeformationresult}, and the HE collimator pointing disparity is shown in Figure  \ref{HEDeformationresult}.
We chose the high temperature working condition in pointing survey mode and the low temperature working condition in the scanning survey mode as the extreme boundary conditions, then we found one typical orbit in each working condition to analyze the pointing deviation angle between each telescope axis and the center star sensor (SS1) axis. The thermal deformation results under the high temperature condition in one typical observation circle show that the maximum pointing deviation angle of HE collimators is 4.46$^\prime$, and the maximum pointing deviation angles of ME and LE detector boxes are respectively 6.28$^\prime$ and 8.66$^\prime$. The HE, ME and LE pointing stability is respectively better than 0.11$^\prime$, 0.72$^\prime$ and 0.69$^\prime$ according to the peak to peak data in Table \ref{tableFEAdeformation}. The thermal deformation results under the low temperature condition in one typical observation circle show that the maximum pointing deviation angle of HE collimators is 6.24$^\prime$, and the maximum pointing deviation angles of ME and LE detector boxes are respectively 9.17$^\prime$ and 12.54$^\prime$. The HE, ME and LE pointing stability is respectively better than 0.027$^\prime$, 0.048$^\prime$ and 0.068$^\prime$ according to the peak to peak data in Table \ref{tableFEAdeformation2}.

The pointing axis deviation of each telescope changes slowly with the on-orbit temperature, so the pointing is relatively stable in a short time such as that in one orbit. 
\begin{table}[htb]
    \centering
    \caption{Pointing deviation angle caused by thermal deformation in high temperature working condition in FEA result.}
    \label{tableFEAdeformation}
    \begin{tabular}{|m{70pt}<{\centering}|m{30pt}<{\centering}|m{30pt}<{\centering}|m{30pt}<{\centering}|m{30pt}<{\centering}|m{30pt}<{\centering}|}
    \hline
        & Average value ($\SI{}{\arcminute}$) & Max value ($\SI{}{\arcminute}$) & Min value ($\SI{}{\arcminute}$) & Peak to Peak ($\SI{}{\arcminute}$) & RMS ($\SI{}{\arcminute}$)\\
    \hline
        HE-collimator-a & 1.027 & 1.048 & 1.008 & 0.039 & 1.027\\
    \hline
        HE-collimator-b & 1.115 & 1.143 & 1.082 & 0.062 & 1.115\\
    \hline
        HE-collimator-c & 3.851 & 3.882 & 3.816 & 0.066 & 3.851\\
    \hline
        HE-collimator-d & 3.846 & 3.879 & 3.812 & 0.067 & 3.847\\
    \hline
        HE-collimator-e & 3.591 & 3.631 & 3.551 & 0.081 & 3.591\\
    \hline
        HE-collimator-f & 3.781 & 3.820 & 3.742 & 0.078 & 3.781\\
    \hline
        HE-collimator-g & 1.058 & 1.078 & 1.033 & 0.045 & 1.058\\
    \hline
        HE-collimator-h & 1.072 & 1.097 & 1.046 & 0.051 & 1.072\\
    \hline
        HE-collimator-i & 3.793 & 3.831 & 3.760 & 0.072 & 3.793\\
    \hline
        HE-collimator-j & 3.908 & 3.940 & 3.887 & 0.053 & 3.908\\
    \hline
        HE-collimator-k & 4.086 & 4.121 & 4.061 & 0.060 & 4.086\\
    \hline
        HE-collimator-m	& 4.214 & 4.263 & 4.173 & 0.090 & 4.214\\
    \hline
        HE-collimator-n	& 1.089 & 1.119 & 1.059 & 0.060 & 1.090\\
    \hline
        HE-collimator-p & 1.037 & 1.056 & 1.016 & 0.040 & 1.037\\
    \hline
        HE-collimator-q & 4.400 & 4.457 & 4.351 & 0.106 & 4.400\\
    \hline
        HE-collimator-r & 4.181 & 4.236 & 4.130 & 0.106 & 4.181\\
    \hline
        HE-collimator-s & 4.265 & 4.306 & 4.226 & 0.080 & 4.265\\
    \hline
        HE-collimator-t & 4.039 & 4.076 & 4.004 & 0.072 & 4.039\\
    \hline
        ME-212a & 5.913 & 6.277 & 5.555 & 0.722 & 5.918\\
    \hline
        ME-211  & 5.295 & 5.467 & 5.120 & 0.346 & 5.296\\
    \hline
        ME-212b & 5.938 & 6.236 & 5.568 & 0.667 & 5.942\\
    \hline
        LE-311a & 8.337 & 8.662 & 7.969 & 0.694 & 8.341 \\
    \hline
        LE-311b & 7.789 & 8.055 & 7.534 & 0.521 & 7.791\\
    \hline
        LE-311c & 8.214 & 8.574 & 7.881 & 0.692 & 8.218\\
    \hline
    \end{tabular}
\end{table}

\begin{table}[htb]
    \centering
    \caption{Pointing deviation angle caused by thermal deformation in low temperature working condition in FEA result.}
    \label{tableFEAdeformation2}
    \begin{threeparttable}  
    
    \begin{tabular}{|m{70pt}<{\centering}|m{30pt}<{\centering}|m{30pt}<{\centering}|m{30pt}<{\centering}|m{30pt}<{\centering}|m{30pt}<{\centering}|}
    \hline
       & Average value ($\SI{}{\arcminute}$) & Max value ($\SI{}{\arcminute}$) & Min value ($\SI{}{\arcminute}$) & Peak to Peak ($\SI{}{\arcminute}$) & RMS ($\SI{}{\arcminute}$)\\
    \hline
        HE-collimator-a & 1.458 & 1.463 & 1.451 & 0.012 & 1.458\\
    \hline
        HE-collimator-b & 1.581 & 1.587 & 1.576 & 0.011 & 1.581\\
    \hline
        HE-collimator-c & 5.470 & 5.481 & 5.459 & 0.022 & 5.470\\
    \hline
        HE-collimator-d & 5.513 & 5.520 & 5.506 & 0.014 & 5.513\\
    \hline
        HE-collimator-e & 5.186 & 5.190 & 5.182 & 0.008 & 5.186\\
    \hline
        HE-collimator-f & 5.487 & 5.496 & 5.478 & 0.018 & 5.487\\
    \hline
        HE-collimator-g & 1.522 & 1.532 & 1.513 & 0.019 & 1.522\\
    \hline
        HE-collimator-h & 1.517 & 1.526 & 1.509 & 0.018 & 1.517\\
    \hline
        HE-collimator-i & 5.502 & 5.515 & 5.489 & 0.026 & 5.502\\
    \hline
        HE-collimator-j & 5.639 & 5.653 & 5.626 & 0.027 & 5.639\\
    \hline
        HE-collimator-k & 5.854 & 5.863 & 5.845 & 0.018 & 5.854\\
    \hline
        HE-collimator-m	& 5.992	& 5.997	& 5.987 & 0.010 & 5.992\\
    \hline
        HE-collimator-n	& 1.539 & 1.542 & 1.535 & 0.007 & 1.539\\
    \hline
        HE-collimator-p & 1.458 & 1.468 & 1.446 & 0.022 & 1.458\\
    \hline
        HE-collimator-q & 6.228 & 6.238 & 6.218 & 0.020 & 6.228\\
    \hline
        HE-collimator-r & 5.903 & 5.916 & 5.889 & 0.027 & 5.903\\
    \hline
        HE-collimator-s & 5.998 & 6.010 & 5.985 & 0.025 & 5.998\\
    \hline
        HE-collimator-t & 5.701 & 5.712 & 5.689 & 0.023 & 5.701\\
    \hline
        ME-212a & 9.147 & 9.168 & 9.120 & 0.048 & 9.147\\
    \hline
        ME-211  & 8.574 & 8.589 & 8.561 & 0.028 & 8.574\\
    \hline
        ME-212b & 9.011 & 9.029 & 8.998 & 0.031 & 9.011\\
    \hline
        LE-311a & 12.495 & 12.532 & 12.464 & 0.068 & 12.494 \\
    \hline
        LE-311b & 11.706 & 11.737 & 11.677 & 0.061 & 11.706 \\
    \hline
        LE-311c & 12.508 & 12.540 & 12.479 & 0.061 & 12.508\\
    \hline
    \end{tabular}
    \end{threeparttable}   
\end{table}

\newpage
 
\subsection{Thermal deformation evaluation based on the variation in relative orientation between star sensors}\label{sec6} 

There are two star sensors set on the telescope reference plane of Insight-HXMT, shown in Figure \ref{fighxmtdistribution}. The main star sensor (SS1) is installed at the center of the reference plane, while the other testing star sensor, the highly precise star sensor (HSS), is installed on the +Y side of the reference plane. Two thermistors TZ140 and TZ141 are separately set close to SS1 and HSS installation positions to monitor the on-orbit temperature, shown in Figure \ref{figPlatedistribution}. Theoretically, both should have the same pointing direction; a slight misalignment could be resulted by thermal deformation. Since the HSS only operated during the first two years after launch, we mainly discuss the data of this period here in this paper.

\begin{figure}[H]%
    \begin{center}
    \includegraphics[width=0.67\textwidth]{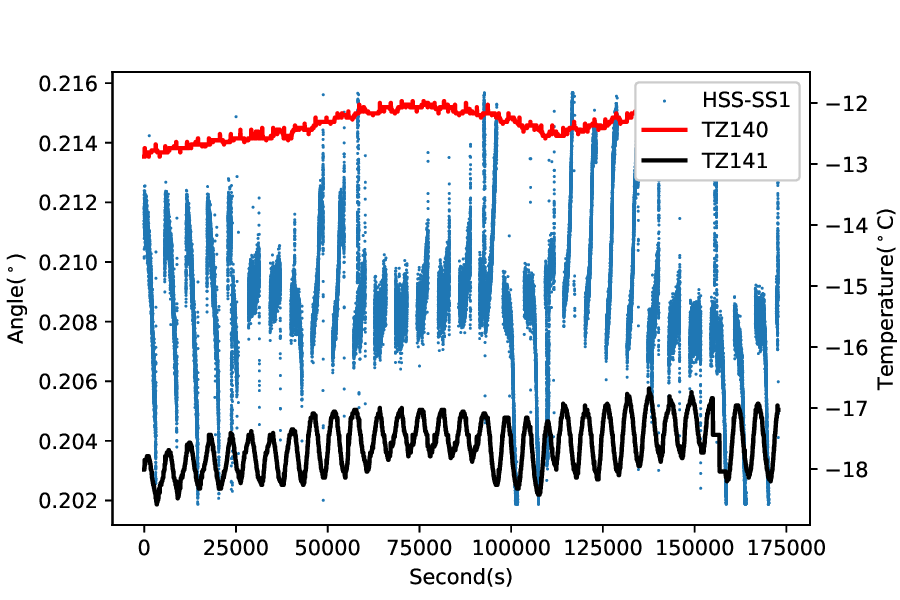}
    \caption{The angle between the two star sensors in typical two days and the corresponding temperature.}
     \label{figSS1andHSSshorttimedata}
     \end{center}
    \footnotesize
    {*}HSS-SS1 blue scatter plot with angle as vertical coordinate represent angle between star sensors HSS and SS1. TZ140 and TZ141 with $\SI{}{\degreeCelsius}$ as vertical coordinate separately represent the temperature close to the SS1 and HSS installation location. 
    \end{figure}

We read the quaternions from SS1 and HSS, converted them to angles, acquired the slight changes on pointing directions of both sensors, and analyzed the relationship between these changes and the temperature data in orbit. Figure \ref{figSS1andHSSshorttimedata} shows how the star sensors' axes pointing and the temperature data varied within a short time, such as between 10th and 11th August 2018. There is a clear relationship between the temperature data and the angle between HSS and SS1 axes as their periods are all about 94 min, which are almost the same as the orbit period. The star sensors were blocked some times, and the data of the angle between the star sensors was discontinuous; the data from the star sensors in these periods are therefore considered invalid.

\begin{figure}[H]%
    \centering
    \includegraphics[width=0.67\textwidth]{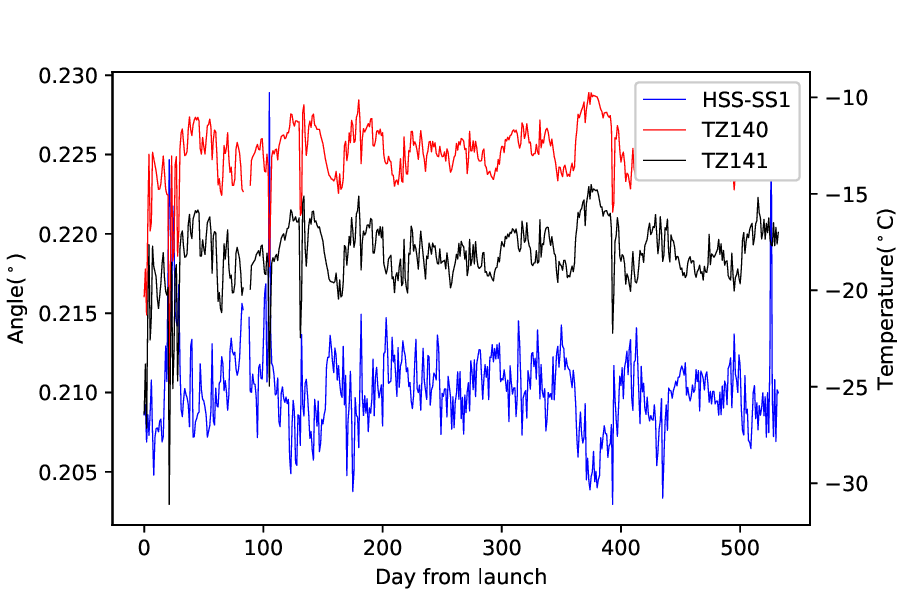}
    \caption{The daily averaged temperature and the angle between star sensors.} \label{figSS1_HSSlongtimedata}
\end{figure}

To analyze the influence of the temperature on the main reference plane in long term, we compared the daily averaged changes of the angle between SS1 and HSS with the thermal data from TZ140 and TZ141, shown in Figure \ref{figSS1_HSSlongtimedata}, where the angle between SS1 and HSS axes is the blue curve marked as “HSS-SS1” in the figure (the value of the angle corresponds to the left ordinate), and the temperature monitoring data is the curve marked as “TZ140” and “TZ141”
(the temperature value corresponds to the right ordinate). It can be seen that variation of the angle is closely correlated with that of the temperatures of TZ140 and TZ141. The lower the temperatures of TZ140 and TZ141, the larger the angle between the SS1 and HSS. 
 
The thermal deformation change of the main structure is about 0.3$^\prime$ in one orbit according to the angle changes between SS1 and HSS.

\begin{figure}[H]%
    \centering
    \includegraphics[width=0.67\textwidth]{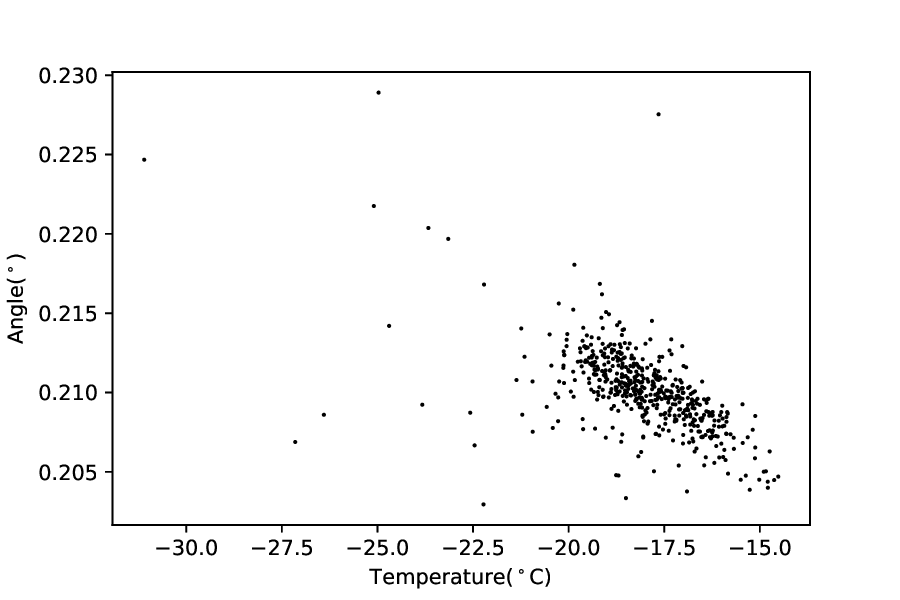}
    \caption{Correlation between the temperature and angle between the axes of SS1 and HSS.} \label{figtemp-sensorangleplot}
\end{figure}

To further understand the relationship between the pointing directions and the temperature data, we created a scatter plot (Figure \ref{figtemp-sensorangleplot}) with the daily averaged temperature changes and the pointing angles between HSS and SS1, and a negative correlation could be seen directly between the temperature and the angle between the star sensors. The Pearson correlation coefficient is -0.54, indicating a strong and negative correlation.

\subsection{Long-term temperature effect on the PSF}\label{sec7} 
\begin{figure}[htb]%
    \centering
    \includegraphics[width=0.8\textwidth]{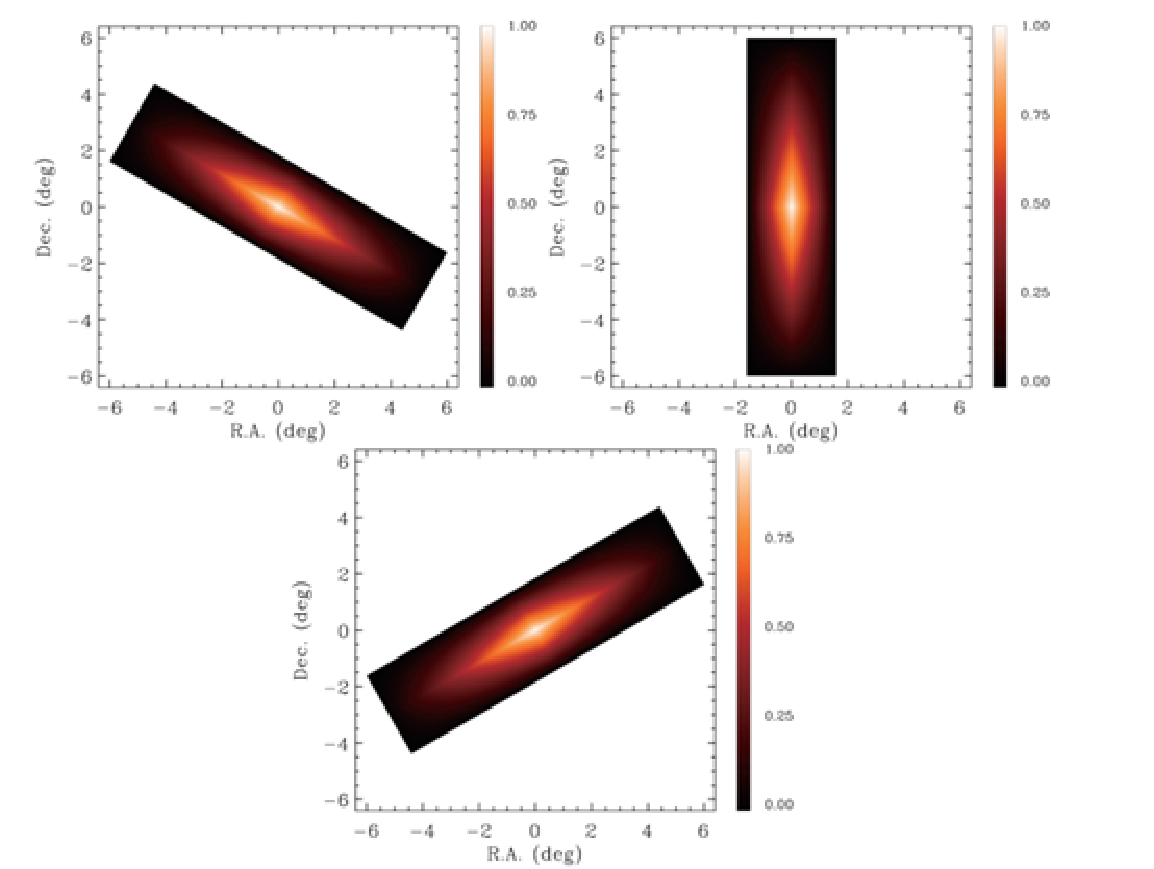}
    \caption{PSFs of three LE boxes with the parameters obtained from the scanning observation to Crab nebula.} \label{figPSFcrab}
\end{figure}

We also analyzed the possible effects of thermal deformation on the PSF.
As described in the paper\cite{2020JHEAp..25...39N}, the real PSF can be modeled by the geometric PSF model with rotation and efficiency corrections. 
The typical PSFs of the three LE boxes are shown in Figure \ref{figPSFcrab}. 
Consequently, the real X-axis angle, rotation angle, and the PSF value can be calculated from the parameters of the corrected PSF model. Here, we defined the pointing deviation ($\Delta P$) as the difference of the pointing angles between the real X-axis angles of the first two years and the fifth year, and the rotation deviation ($\Delta R$) as the difference of the real rotation angles of the first two years and fifth year. In Table \ref{tablePSF}, $\Delta P$ and $\Delta R$ of all LE and ME collimators, as well as six HE collimators are shown respectively. In addition, the minimum and maximum of the PSF difference ($\Delta A$) within FWHM between the first two years and the fifth year are also presented. The maximum deviation is only about 0.9 arcmin, which is much smaller than the pointing control accuracy ($\sim0.1\SI{}\degree$) of the satellite. Therefore, the effect of the thermal distortion on PSF is negligible.

\begin{table}[htb]
    \caption{The deviation of pointing deviation, FOV rotation, and the PSF between the first two years and last year.}\label{tablePSF}
    \centering
         
      \begin{tabular}{*4{c}}\toprule
            & $\Delta P$ ($\times \SI{0.01}{\degree}$) & $\Delta R$ ($\times \SI{0.01}{\degree}$) & $\Delta A$ ($\%$)\tnote{1} \\ \midrule
        HE-collimator-b  &  $1.54\pm0.81$  &  $0.31\pm0.63$  &  (-0.9, 2.8) \\
        HE-collimator-e  &  $0.45\pm0.78$  &  $1.13\pm0.63$  &  (-2.5, 1.0) \\
        HE-collimator-f  &  $1.10\pm0.30$  &  $0.33\pm0.62$  &  (-1.5, 0.9) \\
        HE-collimator-n  &  $1.02\pm0.26$  &  $1.40\pm0.65$  &  (-5.1, 0.9) \\
        HE-collimator-q  &  $0.48\pm0.69$  &  $0.96\pm0.67$  &  (-0.7, 1.9) \\
        HE-collimator-r  &  $1.03\pm0.66$  &  $0.38\pm0.64$  &  (-0.9, 3.9) \\
        \hline
        ME-212a  &  $0.63\pm0.73$  &  $0.11\pm0.85$  &  (-1.1, 0.2) \\
        ME-211   &  $0.81\pm0.80$  &  $1.60\pm0.94$  &  (-0.6, 3.2) \\
        ME-212b  &  $1.28\pm0.67$  &  $0.97\pm0.83$  &  (-1.0, 4.6) \\
        \hline
        LE-311a  &  $0.10\pm0.35$  &  $0.51\pm0.33$  &  (-0.7, 0.1) \\
        LE-311b  &  $1.59\pm0.36$  &  $0.16\pm0.27$  &  (-2.6, 1.4) \\
        LE-311c  &  $1.52\pm0.34$  &  $0.02\pm0.33$  &  (-1.5, 1.8) \\
        \bottomrule
      \end{tabular}
         \begin{tablenotes}    
        \footnotesize              
        \item[1] Minumum and maximum of the PSF difference within FWHM.         
      \end{tablenotes}            
    
  \end{table}

\newpage
\section{Summary}\label{sec8}  
Since the launch more than 5 years ago, the operating temperatures of all detectors of \textit{Insight}-HXMT have been controlled within the designed range and also stable over time. Thanks to the active temperature control, the working temperature of the HE main detectors is precisely controlled within the range of $17.2\sim18.7\SI{}{\degreeCelsius}$ to ensure their stable operation. The temperature variations of all the other detector units with passive temperature control also do not exceed $10\SI{}{\degreeCelsius}$ with an orbit cycle, and there is no tendency for the weekly average temperature of each detector to increase or decrease after 5 years.

The thermal deformation of the main support structure for the HE, ME and LE detectors can be estimated with the variation in relative orientation between the two star sensors installed in its center and +Y side respectively. The analysis results show that the deformation is very small, with only about 0.3 arcmin variation from the edge of the main structure to its center in one orbit, and consistent with the expected value of the FEA results.

The pointing accuracy, derived from the on-orbit PSF calibration scanning observation on Crab in the first two years and the fifth year respectively, also show that the total deformation between all the detector units does not exceed 1 arcmin, which is comparable with the result of star sensor data and is much smaller than the required pointing accuracy of $0.1\SI{}{\degree}$.
As described in Section~\ref{sec7}, the thermal deformation can have an effect on the on-orbit PSF. Therefore, a comprehensive analysis of thermal deformation will be helpful to calibrate the on-orbit PSF more accurately. For example, a more suitable empirical function will be proposed in the future to model the on-orbit PSF of each telescope of \textit{Insight}-HXMT.

\backmatter
\bmhead{Acknowledgments}

This work was based on the data from \textit{Insight}-HXMT mission, a project funded by the China National Space Administration (CNSA) and the Chinese Academy of Sciences (CAS).  We gratefully acknowledge the support from the National Program on Key Research and Development Project (Grant No.2021YFA0718500) from the Ministry of Science and Technology of China (MOST). All authors appreciate the supports from the National Natural Science Foundation of China under Grants 12273043, U1838201, U1838202, U1938102, and U1938108. This work was partially supported by International Partnership Program of Chinese Academy of Sciences (Grant No.113111KYSB20190020).

Appreciate Zeyu Song from IHEP for meticulous translation and revision.

Appreciate Yongping Li from IHEP for helping with star sensor quaternion calculation.

\bmhead{Declaration}
On behalf of all authors, the corresponding author states that there is no conflict of interest.

\bibliography{sn-article}% common bib file

\end{document}